\documentclass{revtex4}

\usepackage[latin1]{inputenc}
\usepackage[english]{babel}
\usepackage{graphicx}
\usepackage{amsmath,amssymb,amsfonts}
\usepackage{multirow}
\usepackage{tikz}
\usepackage{hyperref}
\numberwithin{equation}{section}

\newcommand{\C}{\mathcal{C}}
\newcommand{\ket}[1]{| #1 \rangle}
\newcommand{\bra}[1]{\langle #1 |}

\newcommand{\vide}[2]{\ket{ \omega(#1)}_{#2}}
\newcommand{\lvide}[2]{\ket{ \tilde{\omega}(#1)}_{#2}}

\newcommand{\vacuum}[1]{\ket{\Omega(#1)}}
\newcommand{\excit}[2]{\ket{ \phi(#1)}_{#2}}
\newcommand{\lexcit}[2]{\ket{ \tilde{\phi}(#1)}_{#2}}

\begin{document}
 
\title{Construction of a Coordinate Bethe Ansatz for the asymmetric simple exclusion process with open boundaries}

\author{Damien Simon}
\email{dsimon@thp.uni-koeln.de}
\affiliation{Institut f\"ur Theoretische Physik, Universit\"at zu K\"oln, Z\"ulpicher Strasse 77, 50937 K\"oln, Germany
}

\begin{abstract}
The asymmetric simple exclusion process with open boundaries, which is a very simple model of out-of-equilibrium statistical physics, is known to be integrable. In particular, its spectrum can be described in terms of Bethe roots. The large deviation function of the current can be obtained as well by diagonalizing a modified transition matrix, that is still integrable: the spectrum of this new matrix can be also described in terms of Bethe roots for special values of the parameters. However, due to the algebraic framework used to write the Bethe equations in the previous works, the nature of the excitations and the full structure of the eigenvectors were still unknown. This paper explains why the eigenvectors of the modified transition matrix are physically relevant, gives an explicit expression for the eigenvectors and applies it to the study of atypical currents. It also shows how the \emph{coordinate Bethe Ansatz} developped for the excitations leads to a simple derivation of the Bethe equations and of the validity conditions of this \emph{Ansatz}. All the results obtained by de Gier and Essler are recovered and the approach gives a physical interpretation of the exceptional points The overlap of this approach with other tools such as the matrix Ansatz is also discussed. The method that is presented here may be not specific to the asymmetric exclusion process and may be applied to other models with open boundaries to find similar exceptional points. 
\end{abstract}

\maketitle

\section{Introduction}

Out-of-equilibrium statistical physics has been an active field of research for more than twenty years. Breaking the detailed balance property or imposing currents of particles or energy through a system prevents from applying most of the standard techniques of equilibrium statistical mechanics. The determination of the complete distribution of the currents, beyond the average values, is in general a hard problem since it depends on the whole history followed by the process. Many general symmetry properties \cite{gallavotti,jarzynski,crooks} have been established for a wide class of models and the macroscopic fluctuation theory \cite{bertini,bodineauderrida} makes the computations of some macroscopic quantities possible for diffusive systems. However, it is hard to compare these results with the ones that can be obtained from a microscopic approach since there are only a few examples of exactly solvable models. This comparison becomes necessary for far-from-equilibrium models for which no general macroscopic theory exists yet.

Finding out-of-equilibrium models that can be exactly solved is often less easy than for equilibrium models. One of the reasons, which comes from integrability, is that the out-of-equilibrium character of the system comes from the boundary conditions imposed to the system. Putting the system in contact with reservoirs at different temperatures or different chemical potentials requires to have a description of the interaction between the system and each reservoir. From the point of view of integrability, the diagonalization of the Markov transition matrix of the system requires that \emph{both} the dynamics of the bulk and the one of the contact with the reservoirs have to be integrable, while at equilibrium only the bulk dynamics may be relevant.

\begin{figure}
 \begin{tikzpicture}[scale=0.7]
\draw (0,0) -- (9,0) ;
\draw (0,0) -- (0,0.4) ;
\draw (1,0) -- (1,0.4) ;  
\draw (2,0) -- (2,0.4) ;
\draw (3,0) -- (3,0.4) ;
\draw (4,0) -- (4,0.4) ;
\draw (5,0) -- (5,0.4) ;
\draw (6,0) -- (6,0.4) ;
\draw (7,0) -- (7,0.4) ;
\draw (8,0) -- (8,0.4) ;
\draw (9,0) -- (9,0.4) ;
\draw[->,thick] (-0.4,0.9) arc (180:0:0.4) ; \node at (0.,1.5) [] {$\alpha$};
\draw[->,thick] (0.4,-0.1) arc (0:-180:0.4) ; \node at (0.,-0.8) [] {$\gamma$};
\draw  (2.5,0.5) circle (0.3) [fill,circle] {};
\draw  (4.5,0.5) circle (0.3) [fill,circle] {};
\draw  (5.5,0.5) circle (0.3) [fill,circle] {};
\draw  (8.5,0.5) circle (0.3) [fill,circle] {};
\draw[->,thick] (2.4,0.9) arc (0:180:0.4); \node at (2.,1.5) [] {$q$};
\draw[->,thick] (2.6,0.9) arc (180:0:0.4); \node at (3.,1.5) [] {$p$};
\draw[->,thick] (4.4,0.9) arc (0:180:0.4); \node at (4.,1.5) [] {$q$};
\draw[->,thick] (5.6,0.9) arc (180:0:0.4); \node at (6.,1.5) [] {$p$};
\draw[->,thick] (8.4,0.9) arc (0:180:0.4); \node at (8.,1.5) [] {$q$};
\draw[->,thick] (8.6,0.9) arc (180:0:0.4) ; \node at (9.,1.5) [] {$\beta$};
\draw[->,thick] (9.4,-0.1) arc (0:-180:0.4) ; \node at (9.,-0.8) [] {$\delta$};
 \end{tikzpicture}
 \caption{Transition rates in the asymmetric exclusion process for a system of size $L$.}
 \label{fig:schemaASEP}
\end{figure}
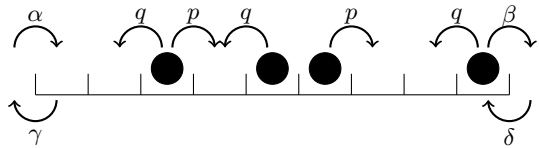

The asymmetric simple exclusion process (ASEP) is one of the simplest out-of-equilibrium models that is integrable. The ASEP describes hard-core particles that diffuse on a one-dimensional lattice of $L$ sites. There is at most one particle per site. Each particle tries to jump to the right (resp. left) neighboring site with rate $p$ (resp.~$q$): if the target site is empty then the jump is performed, if the site is occupied then the particle remains on the initial site (exclusion effect, see fig. \ref{fig:schemaASEP}). The first and the last sites are in contact with reservoirs of particles that can add or remove particles in the system so that they tend to impose their own densities in the system. If the two densities are different and/or if the bias $p-q$ is non-zero, then a current of particles flows through the system from one reservoir to the other.

Many properties of this one-dimensional model of particle transport are now known. A first breakthrough was the introduction of the matrix ansatz \cite{dehp} to study the stationary measure of the ASEP with open boundaries. Although this technique has been fruitfully generalized to many other systems \cite{evansreview}, it remains confined to the study of stationary measures and does not give any information about the relaxation times of the system nor the distribution of the current that flows through the system. The study of finite dimensional representations of the matrix algebra \cite{esslerrittenberg,mallicksandow} and their interpretation in terms of shocks \cite{schuetzjafarpour,jafarpour,belitskyschuetz} give a new physical point of view on the matrix ansatz and is related to the approach I describe below.

Besides the emergence of numerical methods to study the large deviation function of the current in the system \cite{kurchanpeliti,tailleurlecomte,vanderzande}, a second breakthrough has been obtained in \cite{degieressler1,degieressler2,degieressler3}. The Markov transition matrix of the ASEP with reservoirs can be related to the Hamiltonian of the XXZ spin chain with non-hermitian complex boundary fields and to the Temperley-Lieb algebra and many results obtained for the XXZ spin chain \cite{cao,yangzhang,nepomechieravanini,murgannepomechie1,murgannepomechie2,baseilhackoizumi,galleas} or the Temperley-Lieb algebra \cite{degierpyatov,degiernichols} can be extended to the ASEP. A major inconvenient of these methods is that they are often valid only at exceptional points in the parameter space and the validity condition did not have yet any simple physical interpretation in the case of the ASEP. Moreover these methods do not give the structure of the eigenvectors of the Markov transition matrix and thus prevent the computation of correlation functions. This paper presents a simple construction of the eigenvectors based on simple remarks on the expected properties of vacuum states and excitations. These eigenvectors take the form of a coordinate Bethe Ansatz.

This paper is organized as follows: I introduce in section \ref{sec:generalproperties} general properties of the ASEP and show how the current distribution can be obtained and how the coordinate Bethe Ansatz works for periodic boundary conditions. In section \ref{subsec:physrelevance}, I explain the physical interpretation of the eigenvectors build in the next sections. Section \ref{sec:bulkintegrability} is devoted to the study of the bulk dynamics and I show how to construct relevant vacuum states and excitations in presence of two different reservoirs at each end of the lattice. The structure of the excitations in part of the spectrum is similar to the shock structure described in \cite{schuetzjafarpour,jafarpour,belitskyschuetz}, to whom I refer for a physical discussion of the shocks, but I will not focus on it since the construction may be valid for many other systems with different interpretations of the excitations. Section \ref{sec:boundaryintegrability} focuses on the dynamics of an excitation that arrives on a boundary : the Bethe equations \cite{degieressler2} are derived and the interpretation of the boundary integrability is discussed. It also contains the complete formula for the proposed \emph{coordinate Bethe Ansatz}. The last section \ref{sec:application} is an application of section \ref{sec:boundaryintegrability} and shows how the construction of these eigenvectors can be used to study and interpret the dynamics of the system conditioned to produce an atypical current as introduced in section \ref{subsec:physrelevance}.

\section{General properties}
\label{sec:generalproperties}
	\subsection{Definition of the process and generating function of the current}
The asymmetric simple exclusion process corresponds to particles that diffuse on a lattice. We consider here the case of a one-dimensional lattice of length $L$. There is at most one particle per site. If the target site is empty, each particle can jump to the neighbouring site on the right with probability $p dt$ and to the one on the left with probability $q dt$ during an infinitesimal time $dt$. If the target site is occupied, the particle does not attempt to jump. An isolated particle performs thus a random walk. Through the bond between the site $i$ and $i+1$, the transition matrix in the basis of configurations $(0_i0_{i+1},0_i1_{i+1},1_i0_{i+1},1_i1_{i+1})$ is thus given by~:
\begin{equation}
\label{eq:wi}
 w_{i,i+1} =
\begin{pmatrix}
 0 & 0 & 0 & 0 \\
 0 & -q & p & 0 \\
 0 & q & -p & 0 \\
 0 & 0 & 0 & 0
\end{pmatrix}_{i,i+1}.
\end{equation}
One can notice that the number of particles is conserved by this bulk dynamics. In case of periodic boundary conditions, the site $L$ is connected to the site $1$ by a similar local matrix $w_{L,1}$. The diagonalization of the Markov transition matrix using the coordinate Bethe ansatz is presented below in section \ref{subsec:closed:betheansatz}.

For open boundaries, each reservoir tends to impose its own density $\rho_a$ or $\rho_b$ on the neighboring site. The interactions with the reservoirs are described by boundary operators $B_1$ and $B_L$ acting on the sites $1$ and $L$. If site $1$ is occupied, the particle is removed with a rate $\gamma$; if it is empty, a particle is injected with rate $\alpha$. In the same way, a particle is injected by the second reservoir on site $L$ with rate $\delta$ and is removed with rate $\beta$ (see fig.~\ref{fig:schemaASEP}). The two densities the reservoirs try to impose are thus given by $\rho_a=\alpha/(\alpha+\gamma)$ and $\rho_b=\delta/(\beta+\delta)$. In the bases $(0_1,1_1)$ and $(0_L, 1_L)$, the operators $B_1$ and $B_L$ are described by the two matrices:
\begin{subequations}
\label{eq:boundary}
\begin{eqnarray}
 B_1 &=& \begin{pmatrix} -\alpha & \gamma \\ \alpha & -\gamma \end{pmatrix}_1, \\
 B_L &=& \begin{pmatrix} -\delta & \beta \\ \delta & -\beta  \end{pmatrix}_L.
\end{eqnarray}
\end{subequations}
The Markov transition matrix for the open chain is thus given by:
\begin{eqnarray}
 W &=& B_1\otimes I^{\otimes L-1} + \sum_{i=1}^{L-1} I^{\otimes i-1} \otimes w_{i,i+1}\otimes I^{\otimes L-i-1}+  I^{\otimes L-1}\otimes B_L,
\end{eqnarray}
where $I$ is the $2\times 2$ identity matrix. If one introduces the vector $\ket{P_t} = \sum_\mathcal{C} P_t(\mathcal{C}| \C_0) \ket{\mathcal{C}}$ where $P_t(\mathcal{C}| \C_0)$ is the probability that the system is in the configuration $\mathcal{C}$ at time $t$ knowing that the system starts in configuration $\C_0$ and the vectors $\ket{\mathcal{C}}$ form a basis of a $2^L$-dimensional vector space, then one has the evolution
\begin{equation}
 \frac{d}{dt} \ket{P_t} = W\ket{P_t}.
\end{equation}

The diagonalization of $W$ gives the stationary measure and the relaxation times of the ASEP with open boundaries but not the full distribution of the current. Since the system is one-dimensional and the number of particles is conserved in the bulk, the current $Q$ can be measured through any bond and will be defined here as the difference between the integer numbers of particles added on the first site by the left reservoir and of particles removed by the same reservoir. To obtain it, one has to consider the joint probability $P_t(\mathcal{C},Q | \C_0)$ that the system is in configuration $\mathcal{C}$ at time $t$ and that a total current $Q$ has been counted between the initial time and $t$ knowing that the system starts in configuration $\C_0$. This probability evolves as:
\begin{equation}
\label{eq:probacurrent}
 \frac{d}{dt} P_t(\mathcal{C},Q | \C_0) = \sum_{\mathcal{C}'\neq\mathcal{C}} W_{\mathcal{C}\mathcal{C'}} P_t(\mathcal{C}',Q-q_{\mathcal{C}\mathcal{C}'}| \C_0) - \left(\sum_{\mathcal{C}'\neq\mathcal{C}} W_{\mathcal{C'}\mathcal{C}} \right) P_t(\mathcal{C},Q| \C_0),
\end{equation}
where $q_{\mathcal{C}\mathcal{C}'}=\pm 1$ or $0$ is the number of particle exchanged with the left reservoir during the change of configuration $\mathcal{C}'\to\mathcal{C}$. The generating function of $Q$ is defined as $\widehat{P}_t(\mathcal{C},s | \C_0)= \sum_Q e^{s Q} P_t(\mathcal{C},Q | \C_0)$ and satisfies the differential equation:
\begin{equation}
 \frac{d}{dt} \widehat{P}_t(\mathcal{C},s | \C_0) = \sum_{\mathcal{C}'\neq\mathcal{C}} \Big(W_{\mathcal{C}\mathcal{C'}} e^{s q_{\mathcal{C}\mathcal{C}'}}\Big) \widehat{P}_t(\mathcal{C}',s| \C_0) - \left(\sum_{\mathcal{C}'\neq\mathcal{C}} W_{\mathcal{C'}\mathcal{C}} \right) \widehat{P}_t(\mathcal{C},s| \C_0).
\end{equation}
In this linear algebra formalism, the long time behavior of $\widehat{P}_t(\mathcal{C},s | \C_0)$ is obtained by diagonalizing a modified matrix $\widehat{W}$ obtained by replacing the boundary operator $B_1$ by:
\begin{equation}
 \widehat{B}_1 = \begin{pmatrix} -\alpha & \gamma e^{-s} \\ \alpha e^{s} & -\gamma \end{pmatrix}_1.
\end{equation}
This matrix $\widehat{W}$ is not stochastic anymore and the long-time behavior of $\widehat{P}_t(\mathcal{C},s| \C_0)$ is dominated by the first eigenvalue $\mu_1(s)$ of the matrix $\widehat{W}$,
\begin{equation}
 \widehat{P}_t(\mathcal{C},s | \C_0) \propto e^{\mu_1(s) t},
\end{equation}
which corresponds to the large deviation behavior:
\begin{equation}
\label{eq:largedev}
 P_t(\mathcal{C},Q | \C_0) \propto e^{ t f(Q/t) },
\end{equation}
where $f$ and $\mu_1$ are related by the Legendre transformation $\mu_1(s)=\max_j(f(j)+sj)$. Most of the interesting properties are thus contained in the modified matrix $\widehat{W}$ and a way of diagonalizing it with Bethe ansatz methods is presented in the next sections. The integrability of $\widehat{W}$ is related to the Temperley-Lieb algebra formed by the operators $w_{i,i+1}$, $\widehat{B}_1$ and $B_L$:
\begin{subequations}
\label{eq:temperleylieb}
\begin{eqnarray}
& & w_{i,i+1}^2 = -(p+q) w_{i,i+1}, \qquad \widehat{B}_1^2 = -(\alpha+\gamma)\widehat{B}_1, \qquad B_L^2 = -(\beta+\delta) B_L,\\
& & w_{i,i+1}w_{i+1,i+2}w_{i,i+1} =(pq) w_{i,i+1}, \qquad  w_{i+1,i+2}w_{i,i+1}w_{i+1,i+2} =(pq) w_{i+1,i+2}, \\
& & w_{1,2}\widehat{B}_1w_{1,2} = -(\alpha q+p\gamma) w_{1,2},  \qquad w_{L-1,L} B_Lw_{L-1,L} = -(\beta q+p\delta) w_{L-1,L}.
\end{eqnarray}
\end{subequations}
Many interesting properties of this algebra have been used in the context of integrable systems \cite{degierpyatov,degiernichols} and loop models \cite{loopmodels}. For example, the Yang-Baxter equations can be deduced in a straightforward manner from these algebraic identities. The local Hamiltonians of the XXZ spin chain satisfy the same algebra and the state space of each site is also two-dimensional: one goes from the modified matrix $\widehat{W}$ to the XXZ spin chain with complex non-hermitian boundary fields through a simple change of variable discussed in details in \cite{degieressler2}. For this reason, many methods introduced for example in \cite{murgannepomechie1,murgannepomechie2,nepomechieravanini,bajnok} for the XXZ spin chain can be applied successfully \cite{degieressler1,degieressler2,degieressler3} to the ASEP. The next sections will focus however on the point of view and the language of stochastic exclusion processes.

\subsection{The stationary measure and the matrix ansatz}

The first eigenvector of the Markov transition matrix $W$ that has been known \cite{dehp} is the stationary measure (eigenvalue $\Lambda=0$) when one discards any information about the current ($s=0$). A configuration $\mathcal{C}$ of the system is a vector $\mathcal{C}=\tau_1\tau_2\ldots\tau_L$ in $\{0,1\}^L$. The stationary probability of observing a configuration $\mathcal{C}$ is given by the matrix ansatz:
\begin{equation}
\label{eq:matrixansatz}
 P_\text{st}(\tau_1\tau_2\ldots\tau_L) = \frac{1}{Z_L} \big\langle\big\langle \mathcal{W} \big| \prod_{i=1}^L \left( \tau_i D +(1-\tau_i) E \right) \big| \mathcal{V} \big\rangle\big\rangle,
\end{equation}
where the product of the matrices $D$ and $E$ is ordered from $i=1$ on the left to $i=L$ on the right and the vectors $\langle\langle \mathcal{W} |$ and $|\mathcal{V}\rangle\rangle$ are two vectors living in some auxiliary space, which is not the state space of the system. The condition $W\ket{P_\text{st}}=0$ requires that the matrices $D$ and $E$ and the vectors $\langle\langle \mathcal{W} |$ and $|\mathcal{V}\rangle\rangle$ satisfy the algebra \cite{dehp}:
\begin{eqnarray}
 pDE- qED &=& D+E, \\
 \big\langle\big\langle \mathcal{W} \big| \big(\gamma D - \alpha E) &=& \big\langle\big\langle \mathcal{W} \big|, \\
 \big( \beta D - \delta E \big) \big| \mathcal{V} \big\rangle\big\rangle &=& \big| \mathcal{V} \big\rangle\big\rangle.
\end{eqnarray}
For generic values of the parameters, this algebra does not have any finite-dimensional representation, for which the vectors $\langle\langle \mathcal{W} |$ and $|\mathcal{V}\rangle\rangle$ could have a physical interpretation. However, for some submanifolds of the parameter space \cite{mallicksandow,esslerrittenberg}, one can find finite-dimensional matrices $D$ and $E$ satisfying the algebra above which have an interpretation in terms of superposition of shocks diffusing in the systems \cite{jafarpour,schuetzjafarpour,belitskyschuetz}. It would be interesting to understand the link with the eigenvectors obtained through the Bethe ansatz. 

	\subsection{Physical relevance of the eigenvectors for a non-zero parameter $s$}
\label{subsec:physrelevance}

The large deviation function $f(j)$ of the current describes the probability of observing atypical values $Q=j\tau$ of the current during a duration $\tau$. One way to understand these rare events is to characterize the time evolution of the system conditioned on the production of such an atypical current. One is interested in two particular characterizations of this conditioned evolution: the distribution of the \emph{final} state knowing that an atypical current $Q=j\tau$ was observed between $0$ and $\tau$ on one side, and the conditioned transition matrix in the middle of the time interval $[0,\tau]$ on the other side.

The first case is easy and corresponds to the higher order terms of (\ref{eq:largedev}): the probability of observing a configuration $\C$ at $\tau$ conditioned on $j$ can be studied by a saddle-point analysis of the inverse Legendre transform of $\widehat{P}(\C,s | \C_0)$. One has
\begin{equation}
\label{eq:finalcond}
P_\text{final}^{(j)}(\C) = \frac{ P_\tau(\C,j\tau | \C_0)}{\sum_{\C'}P_\tau(\C',j\tau | \C_0)} \underset{\tau\to\infty}{\propto} \langle \C | \mu_1(s) \rangle
\end{equation}
where $\ket{\mu_1(s)}$ is the right eigenvector of $\widehat{W}$ for $s$ such that $f'(j)=-s$. One sees thus that the physical interpretation of the first eigenvector of $\widehat{W}$ is the description of the final state obtained after the observation of an atypical current $j$ such that $s=f'(j)$.

The second case can be obtained by cutting the time interval $[0,\tau]$ into three intervals $[0,\tau_1]$, $[\tau_1,\tau_2]$ and $[\tau_2,\tau]$. One is thus interested in the probability $P_\text{cond}(\C_2, Q_2,\tau_2 | \C_1, Q_1,\tau_1 ; \C_0,\C,Q)$ that the system is in configuration $\C_2$ at $\tau_2$, with an integrated current $Q_2$,  conditioned on the fact that the systems starts in $\C_0$ and ends in $\C$ with a current $Q$ and knowing that the system was in $\C_1$ with a current $Q_1$ at the first intermediate time $\tau_1$. This conditional probability can be written as the ratio of the probabilities of two histories, that can be further decomposed with Markov property:
\begin{eqnarray}
 P_\text{cond}(\C_2, Q_2,\tau_2 | \C_1, Q_1,\tau_1 ; \C_0,\C,Q) &=& \frac{P(\C_0,0,0 \to \C_1,Q_1,\tau_1 \to \C_2,Q_2,\tau_2 \to \C, Q,\tau)}{P(\C_0,0,0 \to \C_1,Q_1,\tau_1 \to \C, Q,\tau)} \nonumber \\
 &=& \frac{  P_{\tau-\tau_2}( \C,Q-Q_2 | \C_2) P_{\tau_2-\tau_1}( \C_2,Q_2-Q_1 | \C_1) P_{\tau_1}( \C_1,Q_1 | \C_0) }{   P_{\tau-\tau_1}( \C,Q-Q_1 | \C_1)P_{\tau_1}( \C_1,Q_1 | \C_0) } \nonumber \\
&=& \frac{  P_{\tau-\tau_2}( \C,Q-Q_2 | \C_2) P_{\tau_2-\tau_1}( \C_2,Q_2-Q_1 | \C_1)  }{   P_{\tau-\tau_1}( \C,Q-Q_1 | \C_1) } \label{eq:probacond2}
\end{eqnarray}
where $P_t(\C,Q|\C')$ is given in (\ref{eq:probacurrent}). The dependence on $\C_0$ disappears because it is ``erased'' by the conditioning on having $\C_1$ at $\tau_1$. If $\tau_2-\tau_1=\Delta \tau$ is fixed and $\tau_1$ and $\tau-\tau_2$ go to infinity, the large deviation behaviour (\ref{eq:largedev}) shows that the dominant contribution to the conditioned probability is obtained when $Q_1$ and $Q_2=Q_1+\Delta Q$ maximize
\begin{equation*}
(\tau-\tau_1-\Delta \tau) f\left( \frac{j\tau-Q_1-\Delta Q}{\tau-\tau_1-\Delta\tau}\right) - (\tau-\tau_1) f\left( \frac{j\tau-Q_1}{\tau-\tau_1}\right).
\end{equation*}
As expected $Q_1$ has to be of order $j \tau_1$, which means that the realization of the atypical current is distributed all over the time interval. A more careful study of the long time behaviour of (\ref{eq:largedev}) as for (\ref{eq:finalcond}) shows that:
\begin{equation}
P_\text{cond}(\C_2, Q_1+\Delta Q,\tau_1+\Delta \tau | \C_1, Q_1,\tau_1 ; \C_0,\C,j\tau) \underset{\tau-\tau_1,\tau_1 \to\infty}{\simeq}
\frac{\langle \mu_1(s) | \C_2\rangle}{\langle \mu_1(s) | \C_1\rangle} e^{-\Delta \tau \mu_1(s) + s \Delta Q} P_{\Delta\tau}( \C_2,\Delta Q | \C_1) 
\end{equation}
where $s$ is given by $f'(j)=-s$ and $\bra{\mu_1(s)}$ is the left eigenvector of the modified matrix $\widehat{W}$. The behaviour of the {r.h.s.} of the previous equation for an infinitesimal time $\Delta \tau \ll 1$ shows that the transition matrix $W_\text{cond}^{(j)}(\C_2,\C_1)$ at an intermediate time conditioned on the production of a current $j$ at a much larger time scale is given by~:
\begin{equation}
\label{eq:conditdynamics}
W_\text{cond}^{(j)}(\C_2,\C_1) = W_{\C_2\C_1} e^{s q_{\C_2\C_1}} \frac{\langle \mu_1(s) | \C_2\rangle}{\langle \mu_1(s) | \C_1\rangle} - \mu_1(s) \delta_{\C_2\C_1}
\end{equation}
One checks that this new matrix has an eigenvalue $\Lambda=0$ that corresponds to the pseudo-stationary state conditioned to produce a current $j$. For $j=j_\text{moy}$, one has $s=0$ and thus $\mu_1(s)=0$ and $\langle \mu_1(s) | \C_2\rangle$ does not depend on $\C_2$: one recovers the stationary measure of $W$ given by the matrix ansatz. 

For $s\neq 0$, the term $\langle \mu_1(s) | \C_2\rangle /\langle \mu_1(s) | \C_1\rangle$ introduces an effective interaction between the particules that tends to enhance or reduce the jumping rates of particles, depending on their environment. A study of this interaction in a simple case is performed in section \ref{sec:application}.

More generally, the eigenvalues of $W_\text{cond}^{(j)}$ are given by $\Lambda_\text{cond}=\Lambda-\mu_1(s)$ where $\Lambda$ is an eigenvalue of $\widehat{W}$ for $s=-f'(j)$. The corresponding eigenvectors are given by:
\begin{equation}
\label{eq:condeigenvectors}
 \langle \C | \Lambda_\text{cond} \rangle = \langle \mu_1(s) | \C \rangle \langle \C | \Lambda \rangle 
\end{equation}
where $\ket{\Lambda}$ is the right eigenvector of $\widehat{W}$ for the eigenvalue $\Lambda$.

This result shows that the characterization of the evolution of the ASEP conditioned to produce an atypical current $j$ involves the right and left eigenvectors of the matrix $\widehat{W}$ for a suitable value of the parameter $s$ conjugated to the current. Moreover, the slowest relaxation times for this conditioned evolution is simply given by $-(\mu_2(s)-\mu_1(s))^{-1}$ where $\mu_2(s)$ is the second eigenvalue of $\widehat{W}$. One notices that the interpretation of the new matrix $W_\text{cond}^{(j)}$ does not seem to have a simple physical counterpart in the quantum XXZ spin chain case. The next sections are devoted to the construction of some of the eigenvectors $\ket{\Lambda}$ by Bethe Ansatz methods.

	\subsection{The coordinate Bethe Ansatz for periodic boundary conditions}
\label{subsec:closed:betheansatz}

The second situation where the eigenvectors of the asymmetric exclusion process are known is the case of periodic boundary conditions. In this case, the number of particles $N$ is constant and the $2^L$-dimensional state space splits into $L+1$ sectors of dimension $L!/(N!(L-N)!)$ where $N$ is the number of particles. Moreover, there is a duality between the sectors of $N$ and $L-N$ particles which corresponds to the particle-hole duality of the ASEP.  Since there is no boundary, the current will be counted on each site, so that one must consider modified operators 
\begin{equation}
\widehat{w}_{i,i+1} =
\begin{pmatrix}
 0 & 0 & 0 & 0 \\
 0 & -q & pe^s & 0 \\
 0 & qe^{-s} & -p & 0 \\
0 & 0 & 0 & 0
\end{pmatrix},
\end{equation}
where $s$ is the parameter conjugated to the current. In the framework of the Bethe ansatz, the state $\ket{00\ldots 0}$ is stable under all the operators $\widehat{w}_{i,i+1}$ since
\begin{equation}
 \widehat{w}_{i,i+1} \ket{0}_i\ket{0}_{i+1} = 0
\end{equation} and it can be used as a vacuum state on which particles (excitations) can be added.
An isolated particle diffuses freely in the system with jumping rates $p$ and $q$:
\begin{eqnarray*}
 \widehat{w}_{i,i+1} \ket{1}_i\ket{0}_{i+1} &=& pe^s \ket{0}_i\ket{1}_{i+1} - p  \ket{1}_i\ket{0}_{i+1}, \\
 \widehat{w}_{i,i+1} \ket{0}_i\ket{1}_{i+1} &=& qe^{-s} \ket{1}_i\ket{0}_{i+1} - q  \ket{0}_i\ket{1}_{i+1}.
\end{eqnarray*}
Thus, in the sector $N=1$, the eigenvectors are plane waves $\sum_{i=1}^L z^i \ket{0_1\ldots0_{i-1}1_i0_{i+1}\ldots0_L}$ with eigenvalues $pe^s/z+qe^{-s} z-(p+q)$, such that the parameter $z$ satisfies $z^L=1$ due to the periodicity of the lattice. The Bethe ansatz for general $N$ consists \cite{gaudin} in assuming that eigenvectors are superpositions of plane waves with parameter $z_k$ for each particle, such that an eigenvector $\ket{\psi(z_1,\ldots,z_N)}$ is given by:
\begin{equation}
\label{eq:closed:cba}
 \ket{\psi(z_1,\ldots,z_N)} = \sum_{\text{$\vec{x}$ ord.}}\,\sum_{\sigma \in \mathcal{S}_N} A_\sigma z_{\sigma(1)}^{x_1}z_{\sigma(2)}^{x_2}\ldots z_{\sigma(N)}^{x_N} \ket{0}_1^{x_1-1}\ket{1}_{x_1}\ket{0}_{x_1+1}^{x_2-1}\ket{1}_{x_2}\ldots \ket{0}_{x_N+1}^{L}.
\end{equation}
The notations are defined as follows: $\vec{x}=(x_1,x_2,\ldots,x_N)$ is a vector ordered such that $1\leq x_1 <x_2<\ldots<x_N\leq L$, the set $\mathcal{S}_N$ is the set of the $N!$ permutations of $N$ elements and $\ket{0}_{j}^{k}$ stands for the tensor product $\ket{0}_j\otimes\ldots\otimes\ket{0}_k$ from site $j$ to site $k$.

By acting with $\widehat{W}$ on configurations with no pair of neighboring particles, one checks that the eigenvalue $\Lambda(z_1,\ldots,z_N)$ is given by the sum:
\begin{equation}
\label{eq:closed:eigenval}
 \Lambda(z_1,\ldots,z_N) = \sum_{k=1}^N \left( \frac{pe^{-s}}{z_k}+ qe^{-s} z_k -p-q\right).
\end{equation}
When two particles are on adjacent sites, the exclusion effect breaks the simple diffusion of the particles and the coefficients $A_\sigma$ have to be tuned so that $\ket{\psi(z_1,\ldots,z_N)} $ is still an eigenvector. For two particles with Bethe coefficients $z_{\sigma(k)}$ and $z_{\sigma(k+1)}$ on sites $i$ and $i+1$, one can allow the particles to exchange their Bethe numbers and, in order to have an eigenvector, one must have:
\begin{equation}
\label{eq:closed:adjacent}
\begin{split}
  \widehat{w}_{i,i+1} &\Big( A_\sigma z_{\sigma(k)}^i z_{\sigma(k+1)}^{i+1} + A_{\sigma\circ\tau_{k,k+1}} z_{\sigma(k+1)}^i z_{\sigma(k)}^{i+1} \Big) \ket{1}_i\ket{1}_{i+1} \\
 = &  \Big[ \left( \frac{p e^s}{z_{\sigma(k+1)}}+qe^{-s} z_{\sigma(k)} -p-q  \right)  A_\sigma z_{\sigma(k)}^i z_{\sigma(k+1)}^{i+1} \\
&+ 
\left( \frac{p e^s}{z_{\sigma(k)}}+qe^{-s} z_{\sigma(k+1)} -p-q  \right)
A_{\sigma\circ\tau_{k,k+1}} z_{\sigma(k+1)}^i z_{\sigma(k)}^{i+1} \Big] \ket{1}_i\ket{1}_{i+1}.
\end{split}
\end{equation}
Since $ \widehat{w}_{i,i+1} \ket{1}_i\ket{1}_{i+1}=0$, one deduces immediately from the previous equation that the amplitudes must satisfy:
\begin{subequations}
\label{eq:closed:ratioamplitudes}
\begin{eqnarray}
\frac{ A_{\sigma\circ\tau_{k,k+1}} }{ A_\sigma } &=& - \frac{a_\text{period}\left(z_{\sigma(k+1)},z_{\sigma(k)}\right)}{ a_\text{period}\left(z_{\sigma(k)},z_{\sigma(k+1)}\right)}, \\
 a_\text{period}(z,z') &=& pe^s + q e^{-s} z z' -(p+q) z.
\end{eqnarray}
\end{subequations}
If one starts from a permutation $\sigma$ and makes the first particle exchange its Bethe root $z_{\sigma(1)}$ with the second one, then makes the second particle exchange its Bethe root with the third one and so on until the last one, one obtains an amplitude $A_{\sigma\circ c_{2\ldots N1}}$ where $c_{2\ldots N1}$ is a cycle of length $L$. Because of the periodic boundary conditions, the $N$-th particle can also exchange its Bethe root with the first particle when they are on the sites $L$ and $L+1 \equiv 1$ and one must recover the initial amplitude $A_\sigma$ up to a factor $z_{\sigma(1)}^L$. The Bethe roots $z_j$ must then satisfy the equation for all $j$ in $\{1,\ldots,n\}$:
\begin{equation}
\label{eq:closed:betheeqs}
 z_j^L = (-1)^{N-1} \prod_{k=1,k\neq j}^N \frac{ a_\text{period}(z_j,z_k )}{a_\text{period}(z_k,z_j)} 
\end{equation}

The resolution of these equations, at least numerically for finite $N$ and $L$, or analytically for $N,L\to \infty$, gives the complete spectrum with (\ref{eq:closed:eigenval}): a detailed study of these equations (\ref{eq:closed:betheeqs}) was performed in \cite{gwaspohn1,gwaspohn2} in relation with noisy Burgers equation and the six-vertex model.  Eq.~(\ref{eq:closed:cba}) gives the eigenvectors and allows one to compute correlation functions (see \cite{izerginkorepin,kitaninemailletI,kitaninemailletII} for the XXZ spin chain).

Equations similar to (\ref{eq:closed:betheeqs}) have been derived for special sets of parameters for the ASEP with \emph{open} boundaries \cite{degieressler1,degieressler2,degieressler3}. In this case however, the number of particles is not conserved at the boundaries and the integer $N$ in (\ref{eq:closed:betheeqs}) is replaced by some integer $k$ that depends on the parameters $\alpha$, $\beta$, $\gamma$, $\delta$, $p$, $q$ and $s$. The procedure followed in \cite{murgannepomechie1,murgannepomechie2,nepomechieravanini} is based on special properties of the transfer matrix and avoids the question of the eigenvectors. The algebraic Bethe ansatz used in \cite{cao,yangzhang} for part of the spectrum of the XXZ spin chain was a first step to the determination of the eigenvectors. However the physical interpretation of the integer $k$ in the context of the ASEP and simple approach such as the coordinate Bethe ansatz was still missing. In the next section, I show how one can determine with a simple procedure the excitations, which replace the particles, and how the integer $k$ emerges in a straightforward way. In section \ref{sec:boundaryintegrability}, I tackle the problem of the scattering of these excitations at the boundary and the derivation of the Bethe equations.

\section{Bulk integrability and description of the excitations}
\label{sec:bulkintegrability}

	\subsection{Product measure and the open chain}
For periodic boundary conditions, the particles $\ket{1}$ can be seen as excitations added on a reference vacuum state $\ket{00\ldots0}$. In this case, the vacuum state is also an eigenvector (ground state) of $\widehat{W}$ for the eigenvalue $\Lambda=0$. For open boundaries at $s=0$ (the matrix $\widehat{W}$ is stochastic), the ground state is given by the matrix ansatz (\ref{eq:matrixansatz}) but, up to our knowledge, it has not been possible to use it as a vacuum state on which excitations may be added. Thus, we introduce a difference between the \emph{ground state} which is the eigenvector whose eigenvalue has an extremal real part and the local \emph{vacuum states} which are product states stable under the bulk operators $w_{i,i+1}$ and are used to separate excitations.

If one looks for a tensor product $\ket{\omega}_i\ket{\omega'}_{i+1}$ different from $\ket{0}_i\ket{0}_i$ and $\ket{1}_i\ket{1}_{i+1}$ such that 
\begin{equation}
\label{eq:localvacuumeq}
 w_{i,i+1} \ket{\omega}_i\ket{\omega'}_{i+1} = 0,
\end{equation}
then one sees that, up to  multiplicative constants, the two vectors must be such that\footnote{in the case of the TASEP, $q=0$ and the contributions containing $q$ must be absorbed in the global multiplicative constants so that to avoid division by $q$.}~:
\begin{eqnarray*}
 \ket{\omega}_i &\propto& \ket{0}_i + C \ket{1}_i, \\
 \ket{\omega'}_{i+1} &\propto& \ket{0}_{i+1} + C(p/q) \ket{1}_{i+1},
\end{eqnarray*}
where $C$ is still a \emph{free} complex number. In the four-dimensional state space of two adjacent sites, it shows that a \emph{third} tensor product satisfying (\ref{eq:localvacuumeq}) exists besides $\ket{00}$ and $\ket{11}$. In the following sections, this number $C$ will be tuned so that the vacuum states also behave well under the boundary operators $\widehat{B}_1$ and $B_L$.

We thus introduce the vectors $\vide{c}{i}$ on site $i$ defined by~:
\begin{equation}
\label{eq:def:vide}
 \vide{c}{i} = \ket{0}_i + c (p/q)^i \ket{1}_i.
\end{equation}
These vectors satisfy by construction $w_{i,i+1} \vide{c}{i}\vide{c}{i+1} = 0$ and it will be useful to define the vacuum state from site $i$ to site $j$ through
\begin{equation}
\label{eq:vacuumdef}
 \vacuum{c}_i^j = \vide{c}{i}\otimes\vide{c}{i+1}\otimes\ldots\otimes \vide{c}{j}.
\end{equation}
One obtains consequently the stability of the vacuum state under the bulk dynamics:
\begin{equation}
\label{eq:vacuum:bulkstability}
 \left(\sum_{i=1}^{L-1} w_{i,i+1} \right) \vacuum{c}_1^L = 0.
\end{equation}

The boundary operators $\widehat{B}_1$ and $B_L$ are still missing in (\ref{eq:vacuum:bulkstability}). If one wants the vacuum state $\vacuum{c}_1^L$ to be an eigenvector of $\widehat{W}$, then $\vacuum{c}_1^L$ must be an eigenvector of both $\widehat{B}_1$ and $B_L$. These two conditions determine independently twice the same coefficient $c$ and give a constraint on the parameters. For each boundary operator, there are two choices for $c$, which correspond to the two possible eigenvectors. The sign of $c(p/q)^i$ is constant all along the chain and thus only two choices remain out of the four possibilities. The condition for the vacuum state $\vacuum{c}_1^L$ to be an eigenvector and the corresponding eigenvalue are thus summarized in the following table:
\begin{equation}
\label{eq:condition:noexcit}
%\begin{center}
 \begin{tabular}{|c|c|c|}
\hline
 Condition & Eigenvalue & Value of $c$ \\
\hline
 $\alpha\beta(p/q)^{L-1} e^s=\gamma\delta$ & $\Lambda=0$ & $c=e^{s} (\alpha q)/(\gamma p)$ \\
 $(p/q)^{L-1} e^s=1$ & $\Lambda=-\alpha-\beta-\gamma-\delta$ & $c=-e^{s}(q/p)$ \\
\hline
\end{tabular}
%\end{center}
\end{equation}
One checks that these conditions are particular cases of the conditions obtained in \cite{degieressler2,degieressler3} for which the number of Bethe roots in one of the two sectors is $0$ ($k=-L/2$ or $L/2-1$ with the notations of \cite{degieressler2,degieressler3}). Thus the state (\ref{eq:vacuumdef}) corresponds to the case where there is no excitation and can play the role of a vacuum state as expected.

In the case of the XXZ spin chain, the idea of finding vacuum states through local rotations already appeared in \cite{baxter,takhtajanfaddeev} and has been used most in the algebraic Bethe Ansatz framework \cite{takhtajanfaddeev,yangzhang}. In the present case of the ASEP, it appears as a locally stationary two-sites state. In the case where $c>0$, the state (\ref{eq:vacuumdef}) can be seen as a density profile with local Bernoulli measures of intensity $c(p/q)^i/(1+c(p/q)^i)$. In the language of shocks developed in \cite{schuetzjafarpour,jafarpour,belitskyschuetz}, it corresponds to $L-1$ consecutive shocks. When $c<0$, this interpretation in terms of density breaks down.

	\subsection{Cutting the product measure and introducing excitations} 
\label{subsec:open:excitations:def}
Excitations can be defined as local perturbations of the product measure (\ref{eq:vacuumdef}) that diffuse freely under the bulk dynamics $\sum_{i=1}^{L-1} w_{i,i+1}$ when they are isolated. Thus, one can look for excitations $\vacuum{c}_{1}^{x-1}\ket{\phi}_x \vacuum{c}_{x+1}^L$ such that the action of $\sum_{i=1}^{L-1} w_{i,i+1}$ on it gives a linear combination of $\vacuum{c}_{1}^{x'-1}\ket{\phi}_{x'} \vacuum{c}_{x'+1}^L$ where $x'=x+\epsilon$ where $\epsilon=0$ or $\pm 1$. However, no solution with this form can be found. 

A way of relaxing one of the constraint is to consider a state $\vacuum{c}_{1}^{x-1}\ket{\phi}_x \vacuum{c'}_{x+1}^L$ where $c'\neq c$. A solution is found if $c'=c(q/p)$. In this case, the excitation $\ket{\phi}_i$ takes the form:
\begin{equation}
\label{eq:excitv1}
 \ket{\phi}_i = \ket{0}_i + \nu (p/q)^i \ket{1}_i,
\end{equation}
and its dynamics is the same as the one of a single particle $\ket{1}_i$ among empty sites~: it jumps on site $i+1$ with rate $p$ and on site $i-1$ with rate $q$. In order to make the computations of section \ref{sec:boundaryintegrability} easier, it is useful to expand $\ket{\phi}_i$ on the two vectors $\vide{c}{i}$ and $\vide{c'}{i}$ (they form a basis\footnote{for $p=q$ (SSEP), the decomposition is not possible and one should better work with the form (\ref{eq:excitv1}).} as soon as $p\neq q$) and replace the parameter $\nu$ which is still free by a parameter $t$, also independent of $i$, such that~:
\begin{equation}
\label{eq:excitv2}
 \excit{t,c,c'}{i} = \left(\sqrt{\frac{q}{p}}\right)^i \Big( t \vide{c}{i} + (1-t) \vide{c'}{i}\Big).
\end{equation}
where, once again, $c'=(q/p)c$. The factor $(\sqrt{q/p})^i$ is also introduced for later convenience such that the Bethe root of an excitation that has an initial Bethe root $z$ and that is reflected becomes $z^{-1}$ without additional factor even with $p\neq q$. The case $t=0$ makes the first vacuum state end at site $i-1$ and the second start at site $i$; the choice $t=1$ makes the first vacuum end at site $i$ and the next one start at site $i+1$. Intermediate values of $t$ give a superposition of both and seem redundant: we will see however in section \ref{sec:boundaryintegrability} that it can be useful to tune the value of $t$.

The action of a single operator $w_{i,i+1}$ on an excitation involves the action of this operator $w_{i,i+1}$ on a tensor product $\vide{c}{i}\vide{c'}{i+1}$ and one verifies that:
\begin{equation}
 w_{i,i+1}\vide{c}{i}\vide{c'}{i+1} = p\vide{c'}{i}\vide{c'}{i+1} + q \vide{c}{i}\vide{c}{i+1} -(p+q)\vide{c}{i}\vide{c'}{i+1}.
\end{equation}
One deduces that an excitation moves according to:
\begin{subequations}
\label{eq:excitation:move} 
\begin{eqnarray}
 w_{i,i+1} \vide{c}{i}\excit{t,c,c'}{i+1} = \sqrt{pq} \excit{t,c,c'}{i}\vide{c'}{i+1}-q \vide{c}{i}\excit{t,c,c'}{i+1} + \vide{c}{i}\ket{v}_{i+1}, \\
w_{i,i+1} \excit{t,c,c'}{i}\vide{c'}{i+1} = \sqrt{pq} \vide{c}{i}\excit{t,c,c'}{i+1}-p \excit{t,c,c'}{i}\vide{c'}{i+1} - \ket{v}_i \vide{c'}{i+1},
\end{eqnarray}
\end{subequations}
where $\ket{v}_i= (\sqrt{q/p})^i(q-p) \ket{0}_i$. The opposite signs of the telescopic terms $\pm\ket{v}_i$ imply that they disappear up to boundary terms under the action of $\sum_{i} w_{i,i+1}$. 

The action of $\sum_{i=1}^{L-1} w_{i,i+1}$ on a state \begin{equation*}\vacuum{c_1}_1^{x_1-1}\excit{t_1,c_1,c_2}{x_1}\vacuum{c_2}_{x_1+1}^{x_2-1}\excit{t_2,c_2,c_3}{x_2}\ldots\vacuum{c_{n+1}}_{x_n+1}^L,\end{equation*}
with $x_k+1<x_{k+1}$ (no excitations on adjacent sites) and $c_{k+1}=(q/p)c_k$, thus makes each excitation at site $x_i$ jump on the sites $x_i\pm i$ or stay on the same site. Thus, the superposition of plane waves 
\begin{equation}
\label{eq:ansatz:nexcits}
 \ket{\psi(z_1,\ldots, z_n)} \propto  \sum_{\text{$\vec{x}$ ord.}} \sum_{\sigma\in\mathcal{S}_n} A_\sigma \left(\prod_{k=1}^n z_{\sigma(k)}^{x_k}\right)  \vacuum{c_1}_1^{x_1-1}\excit{t_1,c_1,c_2}{x_1}\vacuum{c_2}_{x_1+1}^{x_2-1}\excit{t_2,c_2,c_3}{x_2}\ldots\vacuum{c_{n+1}}_{x_n+1}^L
\end{equation}
behaves, \emph{up to boundary terms}, as eigenvectors of the bulk dynamics
\begin{equation}
\label{eq:ansatz:bulk}
 \left(\sum_{i=1}^{L-1} w_{i,i+1}\right) \ket{z_1\ldots z_n} = \Lambda_\text{bulk}(z_1,\ldots,z_n) \ket{z_1,\ldots, z_n} + \text{boundary terms},
\end{equation}
with the eigenvalue given by
\begin{equation}
\label{eq:bulkeigenvalue:def}
 \Lambda_{\text{bulk}}(z_1,\ldots,z_n) = \sum_{k=1}^n \left( \sqrt{pq}\left( \frac{1}{z_k}+z_k\right) -p-q\right) =\sum_{k=1}^n \lambda(z_k),
\end{equation}
as long as the scattering of excitations on adjacent sites satisfies integrability conditions. The discussion of the boundary terms in (\ref{eq:ansatz:bulk}) is the object of section \ref{sec:boundaryintegrability} and the discussion of the scattering of two adjacent excitations is performed in section \ref{subsec:open:scattering} below. 

Several remarks have to be noticed at this stage. First, the vacuum states $\vacuum{c_k}_{x_{k-1}+1}^{x_k-1}$ and $\vacuum{c_{k+1}}_{x_k+1}^{x_{k+1}-1}$ on the left or the right of an excitation are \emph{not} the same since $c_{k+1}=(q/p)c_k \neq c_k$, except for the symmetric simple exclusion process (SSEP). This difference explains why none of the eigenvectors can be used as a vacuum state and why we have introduced a distinction between the ground state and vacuum states at the beginning of section \ref{sec:bulkintegrability}.

The second remark relies on the identity $\vide{c_n}{i} = \vide{c_{n+1}}{i+1}$. For $t=0$ or $t=1$ (for other values of $t$, the excitations as in (\ref{eq:excitv2}) can always be decomposed), one sees that, near the site of the excitations, the states are locally products of two identical Bernoulli measures and one recognizes a similar structure as the one described in \cite{jafarpour,schuetzjafarpour}. The authors of \cite{jafarpour,schuetzjafarpour} describe the states as combinations of shocks that separate Bernoulli product measure; here, on the contrary, the vacuum states are made of shocks separated by excitations that can be thought as Bernoulli product measures. This duality shocks/Bernoulli product measures seems to play a role similar to the particle-hole duality present in the periodic lattice when the number of particles is conserved.

One can now attempt to construct an eigenvector from the state (\ref{eq:ansatz:nexcits}). If one discards the scattering of an excitation on one boundary (see section \ref{sec:boundaryintegrability}) and one requires that the first vacuum state $\vacuum{c_1}_1^{x_1-1}$ and the last one $\vacuum{c_{n+1}}_{x_n+1}^L$ are eigenvectors of the boundary operators $\widehat{B}_1$ and $B_L$, then the results (\ref{eq:condition:noexcit}) have to be replaced for $n$ excitations by:
\begin{equation}
\label{eq:condition:nexcits}
 \begin{tabular}{|c|c|c|c|c|}
\hline
  Name & Condition & Eigenvalue $\Lambda$ & Value of $c_1$ & Value of $c_{n+1}$ \\
\hline
(A) & $\dfrac{\alpha\beta}{\gamma\delta} \left(\dfrac{p}{q}\right)^{L-1-n} e^s = 1$ & $\Lambda_\text{bulk}(z_1,\ldots,z_n)$ & $c_1 = e^s(\alpha q)/(\gamma p)$ & $c_{n+1}=(\delta q^L)/(\beta p^L) $\\
\hline
(B) & $\left(\dfrac{p}{q}\right)^{L-1-n} e^s = 1$ &  $\Lambda_\text{bulk}(z_1,\ldots,z_n)-(\alpha+\beta+\gamma+\delta)$ & $c_1=-e^s (q/p)$ & $ c_{n+1}= - (q^L/p^L)$ \\
\hline
 \end{tabular}
\end{equation}
The conditions presented in this table are exactly the ones obtained in \cite{degieressler1,degieressler2,degieressler3}, for which the spectrum is parameterized by Bethe roots $z_k$. The next sections are devoted to the determination of the Bethe equations, of the value of the $t_i$ contained in the excitations and to the study of the boundary terms in (\ref{eq:ansatz:bulk}). The final form of the coordinate Bethe ansatz is given in section \ref{sec:boundaryintegrability}.

\begin{figure}
\begin{center}
\begin{tikzpicture}[xscale=0.7,yscale=0.6]
\path[->] (0,0) edge (11,0);
\path[->] (0,0) edge (0,7);
\node at (0,0) [anchor=north] {$1$};
\node at (10,0) [anchor=north] {$L$};
\node at (11,0) [anchor=west] {site};
\node at (0,7) [anchor=east] {$\log( c_1(p/q)^{i-m_i})$};
\path[->] (10,0) edge (10,7);
\path[dashed] (0,1) edge (10,6);
\path[ultra thick] (0,1) edge (2,2) ;
\path[ultra thick] (2,2) edge (3,2) ;
\path[ultra thick] (3,2) edge (4.5,2.75);
\path[dashed] (4.5,2.75) edge (10,5.5);
\path[ultra thick] (4.5,2.75) edge (5.5,2.75);
\path[ultra thick] (5.5,2.75) edge (7.5,3.75);
\path[ultra thick] (7.5,3.75) edge (8.5,3.75);
\path[dashed] (7.5,3.75) edge (10,5.);
\path[ultra thick] (8.5,3.75) edge (10.,4.5);
\node at (0,1) [anchor=east] {$\log(\frac{\alpha}{\gamma}e^s)$};
\node at (10,6) [anchor=west] {$\log(\frac{\alpha}{\gamma}(\frac{p}{q})^{L-1}e^s)$};
\node at (10,4.5) [anchor=west] {$\log(\frac{\alpha}{\gamma}(\frac{p}{q})^{L-1-n}e^s)=\log(\frac{\delta}{\beta})$};
\node at (3,0) [anchor=north] {$x_1$};  \path[dotted] (3,0) edge (3,2);
\node at (5.5,0) [anchor=north] {$x_2$};  \path[dotted] (5.5,0) edge (5.5,2.75);
\node at (8.5,0) [anchor=north] {$x_3$};  \path[dotted] (8.5,0) edge (8.5,3.75);
\end{tikzpicture}
\end{center}
 \caption{Schematic plot of the component (in logarithmic scale) along the basis vector $\ket{1}_i$ of each vector in the tensor product  $\vacuum{c_1}_1^{x_1-1}\excit{t,c_1,c_2}{x_1}\vacuum{c_2}_{x_1+1}^{x_2-1}\ldots\vacuum{c_{n+1}}_{x_n+1}^L$ of $L$ vectors for $n=3$: the component at site $i$ is given by $c_1 (p/q)^{i-m_i}$ where $m_i$ counts the number of excitations between sites $1$ and $i$. The picture represents the case $c_i>0$ and $t=0$ with $p>q$.}
\end{figure}

	\subsection{Scattering of two excitations on adjacent sites}
\label{subsec:open:scattering}	
To have an eigenvector (\ref{eq:ansatz:bulk}) of the bulk dynamics up to boundary terms, one must check that the scattering of two excitations on adjacent sites $\excit{t_k,c_k,c_{k+1}}{i}\excit{t_{k+1},c_{k+1},c_{k+2}}{i+1}$ is compatible with the isolated dynamics of the excitations. As for the case of periodic boundary conditions, the amplitudes $A_\sigma$ can be adjusted to satisfy this constraint, as in (\ref{eq:closed:adjacent}). However, from (\ref{eq:excitation:move}), one sees that additional terms $\ket{v}_i$ and $\ket{v}_{i+1}$ have to be introduced. Amplitudes and coefficients $t_k$ also have to be adjusted such that:
\begin{equation}
\label{eq:bulk:collision}
 \begin{split}
w_{i,i+1}& \Big( A_\sigma z_{\sigma(k)}^i z_{\sigma(k+1)}^{i+1} + A_{\sigma\circ\tau_{k,k+1}} z_{\sigma(k+1)}^i z_{\sigma(k)}^{i+1} \Big) \excit{t_k,c_k,c_{k+1}}{i}\excit{t_{k+1},c_{k+1},c_{k+2}}{i+1} \\
%%%%%%%%%%%
=& \Bigg[\left( \sqrt{pq}\left( \frac{1}{z_{\sigma(k+1)}} + z_{\sigma(k)}  \right) -p-q \right) A_\sigma z_{\sigma(k)}^i z_{\sigma(k+1)}^{i+1} \\
%%%%%%%%%%%%%
&+ \left( \sqrt{pq}\left( \frac{1}{z_{\sigma(k)}} + z_{\sigma(k+1)}  \right) -p-q \right) A_{\sigma\circ\tau_{k,k+1}} z_{\sigma(k+1)}^i z_{\sigma(k)}^{i+1} \Bigg] \excit{t_k,c_k,c_{k+1}}{i}\excit{t_{k+1},c_{k+1},c_{k+2}}{i+1}\\
&+\Big[ A_\sigma z_{\sigma(k)}^i z_{\sigma(k+1)}^{i+1} + A_{\sigma\circ\tau_{k,k+1}} z_{\sigma(k+1)}^i z_{\sigma(k)}^{i+1} \Big] \Big(\excit{t_k,c_k,c_{k+1}}{i}\ket{v}_{i+1}-\ket{v}_i\excit{t_{k+1},c_{k+1},c_{k+2}}{i+1} \Big).
 \end{split}
\end{equation}
This system of four linear equations has non trivial solutions if and only if:
\begin{equation}
 t_k= t_{k+1}.
\end{equation}
Thus, all the excitations along the lattice are characterized by the same global parameter $t$, which is still free. Moreover, one checks easily that the two amplitudes have to satisfy the same type of equation as (\ref{eq:closed:ratioamplitudes}) for the periodic lattice, up to an irrelevant different normalization of the $z_k$'s:
\begin{subequations}
 \label{eq:ratioamplitudes}
 \begin{eqnarray}
\frac{ A_{\sigma\circ\tau_{k,k+1}} }{ A_\sigma } &=& - \frac{a\left(z_{\sigma(k+1)},z_{\sigma(k)}\right)}{ a\left(z_{\sigma(k)},z_{\sigma(k+1)}\right)} \\
 a(z,z') &=& \sqrt{pq} + \sqrt{pq} z z' -(p+q) z
 \end{eqnarray}
\end{subequations}

These ratios relate the different amplitudes $A_\sigma$. To obtain the Bethe equations in the periodic geometry, one starts with a given $A_\sigma$, permutes a Bethe root with all the other ones and then uses the periodic boundary conditions to recover the initial amplitude: the consistency conditions give the Bethe equations. In the present case, (\ref{eq:ratioamplitudes}) allows one to permute Bethe roots and make one of them go from the first excitation to the last one. The scattering of an excitation on a boundary and the computation of its reflection coefficient are discussed in the next section.

\section{Scattering of the excitations at the boundaries}
\label{sec:boundaryintegrability}
	\subsection{Reservoirs and integrability}
The integrability of a quantum Hamiltonian or a stochastic transition matrix $W$ can be seen as the knowledge of a non-trivial one-parameter family of matrices $\mathbf{t}(z)$ that commute with each other and contain the matrix $W$. The commutation relations $[\mathbf{t}(z),\mathbf{t}(z')]=0$ imply that they can all be diagonalized in a common basis and thus the matrix $W$ itself also. However, there is no general procedure to construct these eigenvectors. In some cases as the ASEP on the periodic geometry and the XXZ spin chain, the structure of the matrices $\mathbf{t}(z)$ provides creation and annihilation operators and a vacuum state from which one builds the eigenvectors with $n$ particles or excitations from the ones with only $n-1$ particles. 

For periodic systems, the systematic construction of the family $\mathbf{t}(z)$ relies on the algebraic properties of the local matrices $w_{i,i+1}$: one needs to find a family of matrices $R_{i,i+1}(z)$ that contains $w_{i,i+1}$ and satisfies the so-called Yang-Baxter equations. These equations are conditions on the interaction between three bodies (three sites for the ASEP). For the XXZ spin chain or the ASEP, they can be deduced from the Temperley-Lieb algebra (\ref{eq:temperleylieb}) satisfied by the $w_{i,i+1}$. 

For the open chain, the systematic construction of the family $\mathbf{t}(z)$ was originally performed by Sklyanin \cite{sklyanin}. The standard Yang-Baxter equations describe the integrability of the bulk dynamics. The boundary interaction operators $\widehat{B}_1$ and $B_L$ have to be integrable also: a family of matrices $K_1(z)$ (resp. $K_L(z)$) is associated to each boundary site, contains the operator $\widehat{B}_1$ and $B_L$ and must satisfy reflection Yang-Baxter equations involving \emph{both} the $R$ and the $K$ matrices \cite{sklyanin}. In the case of the ASEP, the algebraic construction of $K_1(z)$ and $K_L(z)$ also relies on the Temperley-Lieb algebra (\ref{eq:temperleylieb}).

However, the diagonalization in the open case  of the family $\mathbf{t}(z)$ through a creation-annihilation algebra as for the ASEP with a periodic geometry is not possible in general. A creation-annihilation algebra and a vacuum state was found in the XXZ spin chain with non-diagonal boundary terms only at exceptional points in the parameter space \cite{cao}. These exceptional points are exactly the same as in (\ref{eq:condition:nexcits},\ref{eq:condition:nexcits:left}). Attempts to avoid the creation-annihilation algebra that is not valid outside these points use other algebraic properties of the matrices $\mathbf{t}(u)$ (fusion rules, etc). However, they often lead to equations satisfied directly by the eigenvalues and prevent the construction of the eigenvectors \cite{murgannepomechie1,murgannepomechie2,nepomechieravanini}. Other constructions for the XXZ spin chain have also been developed recently \cite{baseilhackoizumi} and may give also the full spectrum in terms of Bethe roots \cite{galleas}.

The point of view followed here tries to avoid as far as possible the special algebraic properties of the model, so that the approach may be adapted more easily to other integrable models which do not necessarily rely on the Temperley-Lieb algebra. It may also help to identify new types of integrable boundary interactions for models whose bulk dynamics is already known to be integrable.

Section \ref{subsec:firstreflection} is devoted to the scattering of the first (resp. the last) excitation $\excit{t,c_1,c_2}{i}$ (resp. $\excit{t,c_n,c_{n+1}}{i}$) on the left (resp. right) boundary; reflection and transmission coefficients are computed. Section \ref{subsec:open:betheeqs} uses the expression of the reflection coefficients on the boundary to establish the Bethe equations satisfied by the roots $z_k$. The complete expression of the eigenvectors, i.e. the coordinate Bethe Ansatz, is presented in Section \ref{subsec:ansatzvalidity} and its consistency is also checked. Finally, the formalism is extended to left eigenvectors of $\widehat{W}$ as well and the whole spectrum, as obtained in \cite{degieressler1,degieressler2,degieressler3}, is described.

	\subsection{Reflection of a first excitation on a boundary}
\label{subsec:firstreflection}

Sections \ref{subsec:open:excitations:def} and \ref{subsec:open:scattering} have presented how excitations diffuse under the bulk dynamics. This section presents how an excitation gets scattered when it reaches a boundary. Computations are presented in details for the left reservoir but are valid for both boundaries (by changing $\alpha\leftrightarrow \delta$, $\gamma\leftrightarrow \beta$, $p\leftrightarrow q$ and $x\leftrightarrow L+1-x$).

For closed boundaries ($\widehat{B}_1=B_L=0$), the number of particles is conserved. In this case, coordinate Bethe ansatz takes a form similar to (\ref{eq:closed:cba}) except that each Bethe root $z_k$ can appear also with its inverse $z_k^{-1}$ corresponding to a plane wave that propagates in the reverse direction. The amplitudes $A_\sigma$ have to be replaced by amplitudes $A_{\sigma,\vec{\epsilon}}$ where $\vec{\epsilon}=(\epsilon_1,\ldots,\epsilon_N)$ is a vector of the hypercube with $\epsilon_k=\pm 1$ representing the direction of propagation of each Bethe root. When a particle with Bethe root $z_k$ arrives at site $1$ or $L$, the boundary conditions induce a relation between $A_{\sigma,\vec{\epsilon}}$ and $A_{\sigma,r_k\vec{\epsilon}}$ where $r_k\vec{\epsilon}$ is the same vector as $\vec{\epsilon}$ except that the $k$-th component has a flipped sign. The consistency conditions of all the amplitudes give the Bethe equations.

In the case of the ASEP, the boundaries are not closed for the excitations when the parameters take generic values. Tuning the value of $t$ introduced in (\ref{eq:excitv2}) can close a boundary for the excitations  (see below) but not the second one. Since there is no symmetry-related reason to close one or the other by tuning $t$, this parameter will be kept as a generic parameter. 

Since the boundary are not closed for excitations, an excitation that carries a Bethe root $z$ and arrives at the boundary can either be reflected or removed. In the first case, it gives a second plane wave with Bethe root $z^{-1}$ and an amplitude multiplied by a reflection coefficient $R_1(z)$. In the second case, the state of site $1$ is the \emph{second} vacuum state $\vide{c_2}{1}$ with an amplitude multiplied by a transmission coefficient $T_1(z)$. Moreover, in the case of the removal of an excitation, there are only $n-1$ excitations that contribute to the bulk part of the eigenvalue (\ref{eq:bulkeigenvalue:def}): the determination of $T_1(z)$ must take into account that $\widehat{B}_1$ acting the second vacuum state $\vide{c_2}{1}$ must recover the missing contribution $\lambda(z)=\sqrt{pq}(z+1/z)-p-q$ to the bulk part of the eigenvalue. Moreover, depending on the condition chosen in (\ref{eq:condition:nexcits}), the first vacuum state $\vide{c_1}{1}$ is an eigenvector of $\widehat{B}_1$ with eigenvalue $\Lambda_1=0$ or $-\alpha-\gamma$: the action of $\widehat{B}_1$ on the excitation or on the second vacuum state must also recover this contribution $\Lambda_1$.

The boundary operator $\widehat{B}_1$ induces a coupling between the three states $A_1z\excit{t,c_1,c_2}{1}$, which corresponds to a plane wave with Bethe root $z$, $A'_1z^{-1}\excit{t,c_1,c_2}{1}$, which corresponds to a reflected plane wave with Bethe root $z^{-1}$ and $A'_1=R_1(z)A_1$, and the state $A''_1\vide{c_2}{1}$, which corresponds to the second vacuum state with $A''_1=T_1(z)A'_1$. To have an eigenvector, the couplings must satisfy:
\begin{equation}
 \label{eq:boundarycoupling}
\begin{split}
\widehat{B}_1 &\Big[ A_1z\excit{t,c_1,c_2}{1} + A'_1z^{-1}\excit{t,c_1,c_2}{1} + A''_1\vide{c_2}{1}  \Big] \\
= &  \Big( \Lambda_1+\frac{\sqrt{pq}}{z} -q\Big)  A_1z\excit{t,c_1,c_2}{1}  + \Big( \Lambda_1+\sqrt{pq}z -q\Big)  A'_1z^{-1}\excit{t,c_1,c_2}{1} \\
& + \left( A_1z + A'_1z^{-1} \right) \ket{v}_1  + (\Lambda_1+\lambda(z)) A''_1\vide{c_2}{1} 
\end{split}
\end{equation}
The first two terms of the r.h.s. are the contribution to the eigenvalue that is complementary to the one given by the action of $w_{1,2}$. The third term corresponds to the complementary part to the telescopic term $\ket{v}_1$ left by $w_{1,2}$ (see eq.~(\ref{eq:excitation:move})). The fourth term is the contribution to the bulk eigenvalue that the second vacuum state must contain to compensate the disappearance of the first excitation.

To solve (\ref{eq:boundarycoupling}), one must consider the action of $\widehat{B}_1$ on the two vacuum states $\vide{c_1}{1}$ and $\vide{c_2}{1}$. The two cases presented in (\ref{eq:condition:nexcits}) for which $\Lambda_1=0$ or $-\alpha-\gamma$ can be treated simultaneously:
\begin{eqnarray}
 \widehat{B}_1\vide{c_1}{1} &=& \Lambda_1 \vide{c_1}{1} \\
 \widehat{B}_1\vide{c_2}{1} &=& (-\alpha-\gamma-\Lambda_1) \vide{c_2}{1}+ \frac{(p+q)\Lambda_1+(\alpha q+\gamma p)}{p}\vide{c_1}{1}
\end{eqnarray}
Expanding (\ref{eq:boundarycoupling}) in the basis $(\vide{c_1}{1},\vide{c_2}{1})$ gives
\begin{eqnarray*}
 C_1 \Big[ (p+q)\Lambda_1 + (\alpha q + \gamma p) \Big] &=& p\left[ \frac{\sqrt{pq}}{z} t +q (1-t) \right] Az + p \Big[ \sqrt{pq} z t +q(1-t)\Big] A'_1z^{-1} \\
 C_1 \Big[ \lambda(z) + \alpha+\gamma +2\Lambda_1\Big] &=& \Big[ \sqrt{pq}z (1-t) +p t \Big] Az + \Big[ \frac{\sqrt{pq}}{z} (1-t) + pt\Big] A'_1z^{-1} \\
 C_1 &=& \Big( A_1z+A'_1z^{-1} \Big) (1-t) + \sqrt{p/q}A''_1 
\end{eqnarray*}
The resolution of this system gives the value of $A$ and $A'$ as a function of the global normalization constant $C_1$:
\begin{eqnarray}
 \sqrt{pq} \left( z-\frac{1}{z}\right) \left( \sqrt{\frac{p}{q}}\frac{t}{z}+(1-t) \right)  A_1 z&=& \left( \lambda(z) + \alpha+\gamma+2\Lambda_1 - \frac{(p+q)\Lambda_1+(\alpha q +\gamma p)}{\sqrt{pq} z}\right) C_1\\
\sqrt{pq} \left( \frac{1}{z} - z\right) \left( \sqrt{\frac{p}{q}}t z+(1-t) \right) A'_1z^{-1}&=& \left( \lambda(z) + \alpha+\gamma+2\Lambda_1 - \frac{(p+q)\Lambda_1+(\alpha q +\gamma p)}{\sqrt{pq}}z\right) C_1
\end{eqnarray}
Up to a redefinition of the normalization $C_1$, one obtains~:
\begin{eqnarray}
 A_1 z &=& \bar{C}_1 V_1(z) p_1(z) \left( z-\frac{1}{z} \right) \\
 A'_1 z^{-1} &=& \bar{C}_1 V_1(1/z) p_1(1/z) \left( \frac{1}{z} -z\right) \\
 V_1(z) &=& \lambda(z) + \alpha+\gamma+2\Lambda_1 - \frac{(p+q)\Lambda_1+(\alpha q +\gamma p)}{\sqrt{pq} z} \\
 p_1(z) &=&  \sqrt{\frac{p}{q}}t z+(1-t)
\end{eqnarray}
The reflection coefficient of the plane wave with Bethe root $z$ is given finally by:
\begin{equation}
 \label{eq:reflection1}
 R_1(z) = \frac{A'_1}{A_1} = - \frac{z V_1(1/z) p_1(1/z)}{z^{-1} V_1(z) p_1(z)}
\end{equation}
Moreover, the computation of $A''_1$ as a function of $\bar{C}_1$ gives:
\begin{equation}
\label{eq:transmissioncoeff}
\begin{split}
 A''_1 =& \frac{1}{p}\left(z-\frac{1}{z}\right)^2 \Big[ p^2t^2 +(1-t)^2 \big( pq-(p+q)\Lambda_1-(\alpha q+\gamma p)\big) +  pt (1-t)\big(p+q-\alpha-\gamma-2\Lambda_1\big) \Big] \bar{C}_1  \\
 & = -\left(z-\frac{1}{z}\right)^2 t(1-t) V_1\left(-\sqrt{\frac{p}{q}}\frac{t}{1-t} \right) \bar{C}_1\\
\end{split}
\end{equation}
The second representation of the coefficient is ill-defined for $t=0$ or $1$ since $V_1(z)$ diverges as $z\to 0$. The value of $A''_1$ for $t=0$ or $1$ is obtained by taking the limit $t\to 0$ or $1$ in this second expression.

On the second boundary, the three amplitudes have to satisfy an equation similar to (\ref{eq:boundarycoupling}):
\begin{equation}
\begin{split}
B_L &\Big[ A_Lz^L\excit{t,c_n,c_{n+1}}{L} + A'_Lz^{-L}\excit{t,c_n,c_{n+1}}{L} + A''_L\vide{c_{n}}{L}  \Big] \\
= &  \Big( \Lambda_L+\sqrt{pq}z -p\Big)  A_Lz^L\excit{t,c_n,c_{n+1}}{L}  + \Big( \Lambda_1+\frac{\sqrt{pq}}{z} -p\Big)  A'_Lz^{-L}\excit{t,c_n,c_{n+1}}{L} \\
& - \left( A_Lz^L + A'_Lz^{-L} \right) \ket{v}_1  + (\Lambda_L+\lambda(z)) A''_L\vide{c_n}{L}
\end{split} 
\end{equation}
Similar computations as for the first boundary then give:
\begin{eqnarray}
 A_L z^L &=& \bar{C}_L V_L(1/z) p_L(z) \left( z-\frac{1}{z} \right) \\
 A'_L z^{-L} &=& \bar{C}_L V_L(z) p_L(1/z) \left( \frac{1}{z}-z \right) \\
 A''_L &=&  \left(\sqrt{\frac{q}{p}}\right)^L\left( z-\frac{1}{z} \right)^2 \Big( \sqrt{\frac{q}{p}}  t(1-t)\Big) V_L\left( -\sqrt{\frac{q}{p}}\frac{(1-t)}{t}\right) \bar{C}_L \label{eq:transmissionL}\\
 V_L(z) &=& \lambda(z)+ \beta+\delta +2\Lambda_L - \frac{(p+q)\Lambda_L+(\delta p+\beta q)}{\sqrt{pq} z}  \\
 p_L(z) &=& t + (1-t) \sqrt{\frac{q}{p}} \frac{1}{z} 
\end{eqnarray}
From these expressions, computing the reflexion coefficient $R_L(z)=A'_L/A_L$ and the transmission coefficient $T_L(z)=A''_L/A_L$ is easy. The next section \ref{subsec:open:betheeqs} shows how to extract the Bethe equations from these reflection coefficient $R_1(z)$ and $R_L(z)$ and section \ref{subsec:ansatzvalidity} gives the detailed form of the eigenvectors obtained from these reflexion and transmission coefficients.

One can notice that the dependence on $z$ of the transmitted amplitudes (\ref{eq:transmissioncoeff}) and (\ref{eq:transmissionL}) is all contained in the term $(z-1/z)^2$. The functions $V_1$ and $V_L$ depend only on the boundary rates $\alpha$, $\beta$, $\gamma$ and $\delta$ and on the jumping rates $p$ and $q$ but are independent from the parameter $t$ introduced in the definition of the excitations (\ref{eq:excitv2}). On the contrary, the functions $p_1$ and $p_L$ depend only on $t$ and on the jumping rates $p$ and $q$ but are independent from the boundary rates. Thus the reflexion coefficients $R_1(z)$ and $R_L(z)$ have the form of a product of two terms, one that depends on the boundary rates but not on $t$ and one that has the reverse dependence.

One can also remark that the parameter $s$ conjugated to the current is absent from all these expressions and appears only in the eigenvector through the coefficient $c_k$ that characterizes the vacuum states.

	\subsection{Bethe equations}
\label{subsec:open:betheeqs}

If the transmitted terms left when an excitation leaves the system at one boundary are discarded in this subsection, the coordinate Bethe Ansatz for $n$ excitations (when one of the condition (\ref{eq:condition:nexcits}) is satisfied) is given by
\begin{equation}
\label{eq:cba:leveln}
\begin{split}
 \ket{\psi(z_1,\ldots,z_n)} =& 
\sum_{\text{$\vec{x}$ ord.}}\,\sum_{\sigma\in\mathcal{S}_n}\,\sum_{\vec{\epsilon}\in\mathcal{H}_n} A^{(n)}_{\sigma,\vec{\epsilon}} \prod_{k=1}^n z_{\sigma(k)}^{\epsilon_{\sigma(k)} x_k}\\
& \times \vacuum{c_1}_{1}^{x_1-1}\excit{t,c_1,c_2}{x_1}\vacuum{c_2}_{x_1+1}^{x_2-1}\excit{t,c_2,c_3}{x_2}\vacuum{c_3}_{x_2+1}^{x_3-1}\ldots \vacuum{c_{n+1}}_{x_n+1}^L \\
& + \text{terms with $m<n$ excitations}
\end{split}
\end{equation}
where $\mathcal{S}_n$ is the group of permutations of $n$ elements and $\mathcal{H}_n=\{-1,1\}^n$ is the $n$-dimensional hypercube.

If $j=\sigma(1)$ and $\epsilon_j=+1$, one can permute the Bethe root $z_j$ of the first excitation with the second one $z_{\sigma(2)}^{\epsilon_{\sigma(2)}}$ and then with the third one and so on, until it reaches the $n$-th excitation: each time the two amplitudes $A^{(n)}_{\sigma\circ c_{2\ldots k1},\vec{\epsilon}}$ and $A^{(n)}_{\sigma\circ c_{2\ldots (k+1)1},\vec{\epsilon}}$ are related through the ratio (\ref{eq:ratioamplitudes}). The $n$-th excitation, with Bethe root $z_j$ is reflected with a coefficient $R_L(z_j)$ and the Bethe root becomes $z_j^{-1}$. One can permute this Bethe root $z_j^{-1}$ in the reverse order with all the other ones until it reaches again the first one, and each permutation yields a scattering factor (\ref{eq:ratioamplitudes}). A reflection of the first excitation with Bethe root $z_j^{-1}$ on the first boundary gives back the initial $z_j$ up to a reflection coefficient $R_1(z_j^{-1})=R_1(z_j)^{-1}$. At this stage, one must recover the first amplitude $A_{\sigma,\epsilon}$ and the cumulated product of all the scattering factors and the reflexion coefficients must be equal to $1$. From the cycle of transformations
% \begin{center}
% \begin{tikzpicture}
% \matrix (m) [matrix of math nodes, row sep=3em, column sep=3em, text height=1.5ex, text depth=0.25ex]
% { A_{\sigma,\vec{\epsilon}}    &   A_{\sigma\circ\tau_{12},\vec{\epsilon}}   &   A_{\sigma\circ c_{231},\vec{\epsilon}} & \ldots &  A_{\sigma\circ c_{23\ldots n1},\vec{\epsilon}} \\
% A_{\sigma,r_{\sigma(1)}\vec{\epsilon}}    &   A_{\sigma\circ\tau_{12},r_{\sigma(1)}\vec{\epsilon}}   &   A_{\sigma\circ c_{231},r_{\sigma(1)}\vec{\epsilon}} & \ldots &  A_{\sigma\circ c_{23\ldots n1},r_{\sigma(1)}\vec{\epsilon}} \\
% };
% \path[->] (m-1-1) edge node[auto] {(\ref{eq:ratioamplitudes})} (m-1-2);
% \path[->] (m-1-2) edge node[auto] {(\ref{eq:ratioamplitudes})} (m-1-3);
% \path[->] (m-1-3) edge node[auto] {(\ref{eq:ratioamplitudes})} (m-1-4);
% \path[->] (m-1-4) edge node[auto] {(\ref{eq:ratioamplitudes})} (m-1-5);
% \path[->] (m-2-5) edge node[auto] {(\ref{eq:ratioamplitudes})} (m-2-4);
% \path[->] (m-2-4) edge node[auto] {(\ref{eq:ratioamplitudes})} (m-2-3);
% \path[->] (m-2-3) edge node[auto] {(\ref{eq:ratioamplitudes})} (m-2-2);
% \path[->] (m-2-2) edge node[auto] {(\ref{eq:ratioamplitudes})} (m-2-1);
% \path[->] (m-1-5) edge node[auto] {$R_L(z_{\sigma(1)}^{\epsilon_{\sigma(1)}}) $} (m-2-5);
% \path[->] (m-2-1) edge node[auto] {$R_1(z_{\sigma(1)}^{-\epsilon_{\sigma(1)}}) $} (m-1-1);
% \end{tikzpicture}
% \end{center}
\begin{equation*}
\begin{array}{ccccc}
A_{\sigma,\vec{\epsilon}} & \xrightarrow{\text{(\ref{eq:ratioamplitudes})}}  &
A_{\sigma\circ\tau_{12},\vec{\epsilon}} &
 \xrightarrow{\text{(\ref{eq:ratioamplitudes})}}  \ldots  \xrightarrow{\text{(\ref{eq:ratioamplitudes})}} &
A_{\sigma\circ c_{23\ldots n1},\vec{\epsilon}} \\
%%%%%%%%%%%%%%%%%%%%%%%%%%%%%%
  \uparrow & & & & \downarrow \\
%%%%%%%%%%%%%%%%%%%%%%%%%%%%%%
A_{\sigma,r_{\sigma(1)}\vec{\epsilon}} & \xleftarrow{\text{(\ref{eq:ratioamplitudes})}}  &
A_{\sigma\circ\tau_{12},r_{\sigma(1)}\vec{\epsilon}} &
 \xleftarrow{\text{(\ref{eq:ratioamplitudes})}}  \ldots  \xleftarrow{\text{(\ref{eq:ratioamplitudes})}} &
A_{\sigma\circ c_{23\ldots n1},r_{\sigma(1)}\vec{\epsilon}}
\end{array}
\end{equation*}
one obtains the Bethe equations:
\begin{equation}
 \left( (-1)^{n-1} \prod_{k=1,k\neq j}^n \frac{a(z_k,z_j)}{a(z_j,z_k)} \right) R_L(z_j) \left( (-1)^{n-1} \prod_{k=1,k\neq j}^n \frac{a(z_j^{-1},z_k)}{a(z_k,z_j^{-1})} \right) R_1(z_j^{-1})= 1
\end{equation}
The contributions that comes from $p_1(z)$ and $p_L(z)$ cancel and the simplified Bethe equations are independent from the parameter~$t$ as expected:
\begin{equation}
\label{eq:open:betheeqs}
\boxed{
 z_j^{2L} \frac{V_1(z_j)V_L(z_j)}{V_1(z_j^{-1})V_L(z_j^{-1})} = \prod_{k=1,k\neq j}^n \frac{a(z_j,z_k)}{a(z_k,z_j)}\frac{a(z_k,z_j^{-1})}{a(z_j^{-1},z_k)}
}
\end{equation}
This set of equations for $j=1,\ldots,n$ is exactly the one obtained in \cite{degieressler2,degieressler3} for only one part of the spectrum, after the change of variable $z\mapsto (z\sqrt{q/p}-1)/(\sqrt{q/p}-z)$. If the first condition in (\ref{eq:condition:nexcits}) is satisfied, one has $\Lambda_1=\Lambda_L=0$ and one recovers the second part of the spectrum in \cite{degieressler2} (eq.~(2.13)). On the contrary, if the second condition in  (\ref{eq:condition:nexcits}) is fulfilled, one has $\Lambda_1=-\alpha-\gamma$ and $\Lambda_L=-\beta-\delta$ and one recovers the first part of the spectrum in \cite{degieressler2} (eq.~(2.10)). I explain in section \ref{subsec:lefteigen} how to obtain the second part of the spectrum in both cases by using the same approach on left eigenvectors of the transition matrix $\widehat{W}$.

One can check that the equations (\ref{eq:open:betheeqs}) have the expected symmetries. For example, changing a $z_k$ to $z_k^{-1}$ corresponds to exchanging the roles of a plane wave and its reflected partner and should not change the Bethe equations. One can check indeed that the set of equations (\ref{eq:open:betheeqs}) is indeed invariant under the ``gauge'' transformation $z_k \to z_k^{\epsilon_k}$ where the signs $\epsilon_k=\pm 1$ are independent.

	\subsection{Complete coordinate Bethe Ansatz and validity condition}
\label{subsec:ansatzvalidity}

Eq.~(\ref{eq:cba:leveln}) gives the Bethe equations by considering that the $n$ excitations remain in the system up to reflexion coefficients at the boundaries. However section \ref{subsec:firstreflection} has shown that for generic values of $t$ one must also consider the case where an excitation disappears at the boundary. These terms have to be added to the coordinate Bethe Ansatz to have correct eigenvectors.

It is easy to see under which condition the boundaries are closed for the excitations: the two coefficients (\ref{eq:transmissioncoeff}) and (\ref{eq:transmissionL}) must vanish simultaneously. Introducing the reduced variable $u= -\sqrt{p/q} t/(1-t)$ gives the condition:
\begin{equation}
 V_1(u) = V_L(1/u)=0 
\end{equation}
It is always possible to choose $t$ and thus $u$ so that the first term or the second term vanishes. However, they vanish simultaneously only under the additional assumption on the parameters that $V_1(z)$ and $V_L(1/z)$ have a common zero $u^*$. For $u=u^*$, there is no additional term in (\ref{eq:cba:leveln}). 

For generic values of $t$, the additional terms in (\ref{eq:cba:leveln}) contains only $n-1$ excitations that are moving in the bulk. In order to make computations easier, the free parameter $t$ can be tuned to the value $t=t_L^*$ such that (\ref{eq:transmissionL}) vanishes and excitations can leave the system only through the left reservoir. After the disappearance of $\excit{t,c_1,c_2}{i}$, the second excitation $\excit{t,c_2,c_3}{i}$ can reach also the left boundary. To study its scattering, let us start with $n$ excitations as in (\ref{eq:cba:leveln}) and consider for simplicity the components $\sigma(j)=j$ for $j\geq 3$ and $\epsilon_j=+1$ for $j\geq 3$, i.e. $\sigma=\text{id}$ (identity) or $\sigma=\tau_{12}$ (transposition of the the first two). There are 8 components corresponding to the possible permutations, $\epsilon_1=\pm 1$ and $\epsilon_2=\pm 1$. Notations can be shortened by introducing the eight amplitudes $A_{12}^{\epsilon_1\epsilon_2}= A^{(n)}_{\text{id},\epsilon_1\epsilon_2+\ldots+}$ and $A_{21}^{\epsilon_1\epsilon_2}= A^{(n)}_{\tau_{12},\epsilon_1\epsilon_2+\ldots+}$.

Amplitudes $A_{12}^{\epsilon_1\epsilon_2}$ and $A_{21}^{\epsilon_1\epsilon_2}$ are coupled by (\ref{eq:ratioamplitudes}). Moreover $A_{12}^{(+)\epsilon_2}$ and $A_{12}^{(-)\epsilon_2}$ are coupled by the reflexion coefficient $R_1(z_1)$ and both are coupled with the component, with amplitude $A_{[1]2}^{\epsilon_2}$, in which $\excit{t,c_1,c_2}{i}$ has left the system with the root $z_1$. In the same manner $A_{21}^{\epsilon_1(+)}$ and $A_{21}^{\epsilon_1(-)}$ are coupled with $R_1(z_2)$ and are also coupled with the component, with amplitude $A_{[2]1}^{\epsilon_1}$, corresponding to $n-1$ excitations without $z_2$. Thus the four amplitudes $A_{[1]2}^{\epsilon_2}$ and $A_{[2]1}^{\epsilon_1}$ are proportional to a global normalization constant $K$ with:
\begin{eqnarray}
A_{[1]2}^{\epsilon_2} z_2^{\epsilon_2}&=& \epsilon_2 V_1(z_2^{\epsilon_2})p_1(z_2^{\epsilon_2}) a(z_1,z_2^{\epsilon_2}) a(z_1^{-1},z_2^{\epsilon_2}) z_2^{-\epsilon_2} \left( z_1 -\frac{1}{z_1} \right) K \\
A_{[2]1}^{\epsilon_1} z_1^{\epsilon_1}&=& -\epsilon_1 V_1(z_1^{\epsilon_1})p_1(z_1^{\epsilon_1}) a(z_2,z_1^{\epsilon_1}) a(z_2^{-1},z_1^{\epsilon_1}) z_1^{-\epsilon_1} \left( z_2 -\frac{1}{z_2} \right)K
\end{eqnarray}

If one removes then the second excitation, then the four amplitudes $A_{[1]2}^{\epsilon_2}$ and $A_{[2]1}^{\epsilon_1}$ are coupled to the same state, with amplitude $A_{[12]}$, which contains only $n-2$ excitations with Bethe roots $z_3$,\ldots,$z_n$. The coupling of these five amplitudes is of the same type as in (\ref{eq:boundarycoupling}) except that one must recover the contribution of the Bethe root that has already disappeared from the system at the first stage:
\begin{equation}
\label{eq:boundarycoupling:level2}
 \begin{split}
  \widehat{B}_1&  \Big[ \Big( A_{[2]1}^{+}z_1 + A_{[2]1}^{-}z_1^{-1} + A_{[1]2}^{+}z_2 + A_{[1]2}^{-}z_2^{-1} \Big)\excit{t,c_2,c_3}{1} + A_{[12]}\vide{c_3}{1} \Big] \\
=& \Big[\Big(\Lambda_1 + \lambda(z_2) + \frac{\sqrt{pq}}{z_1} - q \Big) A_{[2]1}^{+}z_1  + \Big(\Lambda_1 + \lambda(z_2) + \sqrt{pq}z_1 - q \Big) A_{[2]1}^{-}z_1^{-1} \\
& + \Big(\Lambda_1 + \lambda(z_1) + \frac{\sqrt{pq}}{z_2} - q \Big) A_{[1]2}^{+}z_2  + \Big(\Lambda_1 + \lambda(z_1) + \sqrt{pq}z_2 - q \Big) A_{[1]2}^{-}z_2^{-1} \Big] \excit{t,c_2,c_3}{1}\\
& + \Big( A_{[2]1}^{+}z_1 + A_{[2]1}^{-}z_1^{-1} + A_{[1]2}^{+}z_2 + A_{[1]2}^{-}z_2^{-1} \Big) \ket{v}_1 + \Big( \Lambda_1+\lambda(z_1)+\lambda(z_2) \Big) A_{[12]}\vide{c_3}{1}
 \end{split}
\end{equation}

The integrability of the boundary interactions with the reservoirs appears only at this point. Indeed, Eq.~(\ref{eq:boundarycoupling}) can be written as three two-dimensional vectors, with three unknown amplitudes $A_1$, $A_1'$ and $A_1''$, whose sum is $0$:
\begin{equation}
 A_1 \ket{a} + A_1' \ket{a'} + A_1'' \ket{a''} = 0
\end{equation}
In the generic case, $\ket{a'}$ and $\ket{a''}$ are independent and it is always possible to find $A_1'$ and $A_1''$ to satisfy this condition. On the contrary, (\ref{eq:boundarycoupling:level2}) can be written as a sum of \emph{five} two-dimensional vectors that should give $0$ but four amplitudes are already fixed (up to an overall normalization constant) by the previous step. The sum can be written formally:
\begin{equation}
\label{eq:compatibility:level2}
 \Big( A_{[2]1}^{+} \ket{b_{21}^+}   +A_{[2]1}^{-} \ket{b_{21}^-}  +  A_{[1]2}^{+}  \ket{b_{12}^+} +A_{[1]2}^{-} \ket{b_{12}^-}  \Big) + A_{[12]} \ket{b} = 0
\end{equation}
where the vector formed by the first four vectors is fixed by the previous step. A solution for $A_{[12]}$ exists if and only if $\ket{b}$ is proportional to the sum of the four other vectors. If it is not the case, the Ansatz breaks down and the Bethe equations (\ref{eq:open:betheeqs}) are meaningless. For the boundary operator $\widehat{B}_1$ and $B_L$ defined in (\ref{eq:boundary}), a lengthy computation shows that the sum of the first four vector is proportional to $\ket{b}$ and a solution exists indeed for $A_{[12]}$. 

For boundary operators, the integrability condition on the scattering of two excitations in the bulk (\ref{eq:bulk:collision}) is replaced by a condition on the transmission coefficients of the first excitations $z_1$ and $z_2$: eq.~(\ref{eq:boundarycoupling:level2}) states that they must be compatible with the reflection coefficients of the second excitation with the same two Bethe roots. The same procedure holds for the removal of the third excitation and, from integrability, one expects it be true until the removal of the last excitation $\excit{t,c_n,c_{n+1}}{1}$ although showing it rigorously would require more efforts.

Thus for $t=t_L^*$ such that the transmission coefficient (\ref{eq:transmissionL}) at the right boundary is $0$, the complete coordinate Bethe ansatz can be written as:
\begin{equation}
 \label{eq:completeCBA}
\begin{split}
\ket{\psi(z_1,\ldots,z_n)} &= 
\sum_{m=0}^{n} \;\sum_{\text{$\vec{x}^{(n-m)}$ ord.}} \;\sum_{\sigma\in\mathcal{S}_{n-m}}\;\sum_{\vec{\epsilon}\in\mathcal{H}_{n-m}} A^{(n-m)}_{\sigma,\vec{\epsilon}} \prod_{k=m+1}^n z_{\sigma(k)}^{\epsilon_{\sigma(k)} x_k}\\
%%%%%%%%%
& \times \vacuum{c_{m+1}}_{1}^{x_{m+1}-1}\excit{t,c_{m+1},c_{m+2}}{x_{m+1}}\vacuum{c_{m+2}}_{x_{m+1}}^{x_{m+2}-1}\excit{t,c_{m+2},c_{m+3}}{x_{m+2}}\ldots \vacuum{c_{n+1}}_{x_n+1}^L \\ 
\end{split}
\end{equation}
where $m$ counts the number of excitations that have been removed at the left boundary. The $(n-m)$-dimensional vector $\vec{x}^{(n-m)}=(x_{m+1},\ldots,x_n)$ with $x_k<x_{k+1}$ give the position of the $n-m$ excitations that remain in the bulk. As before, the sets $\mathcal{S}_{n-m}$ and $\mathcal{H}_{n-m}$ are the group of permutations of $n-m$ objects and the hypercube $\{-1,1\}^{n-m}$.

If $t$ was chosen to be equal to $t_1^*$ so that the left boundary is closed instead of the right one, the coordinate Bethe ansatz (\ref{eq:completeCBA}) would have been a combination of tensor product vectors starting with the vacuum $\vacuum{c_1}_1$ and ending with the vacuum $\vacuum{c_{n-m+1}}_L$. Both expressions should be equal up to a normalization constant.

	\subsection{Second part of the spectrum and left eigenvectors}
\label{subsec:lefteigen}

In section \ref{subsec:open:betheeqs}, only one part of the spectrum described in \cite{degieressler2,degieressler3} is obtained with the ansatz (\ref{eq:cba:leveln}). Up to now, it has not been possible to write down the ansatz that corresponds to the right eigenvectors in the second part of the spectrum. However it is possible to obtain the Bethe equations for this second part by considering the left eigenvectors of the matrix $\widehat{W}$. To simplify the notations, we will write them as right eigenvectors of the transposed matrix $\widehat{W}^T$ and the change $\ket{0}\mapsto\bra{0}$ and $\ket{1}\mapsto\bra{1}$ allows one to go easily from one to the other. Up to a change of basis, $\widehat{W}^T$ with a given value of $s$ is the same as $\widehat{W}$ with a value $s'= -\eta-s$ where $e^\eta=(\alpha\beta)/(\gamma\delta)(p/q)^{L-1}$. This symmetry is known as the Gallavotti-Cohen symmetry and $\widehat{W}^T$ is related to the time-reversed properties of the exclusion process. 

From the definition
\begin{equation}
\label{eq:vide:left}
 \lvide{\tilde{c}}{i} = \ket{0}_i + \tilde{c} \ket{1}_i,
\end{equation}
one checks easily that 
\begin{equation}
   w^T_{i,i+1} \Big( \lvide{\tilde{c}}{i}\otimes\lvide{\tilde{c}}{i+1} \Big)= 0
\end{equation}
and the vectors $\lvide{\tilde{c}}{i}$ can be used to build a vacuum state. Moreover excitations defined by
\begin{equation}
\label{eq:excit:left}
 \lexcit{\tilde{t},\tilde{c},\tilde{c}'}{i} = \left(\sqrt{\frac{p}{q}}\right)^i \Big( \tilde{t}\lvide{\tilde{c}}{i}+(1-\tilde{t})\lvide{\tilde{c}'}{i}\Big),
\end{equation}
where $\tilde{c}'=(q/p)\tilde{c}$,
move with the same dynamics as in (\ref{eq:excitation:move}): the jumping rates are the same and the telescopic term $\ket{v}_i$ has the same expression. A state with $n'$ excitations inside the bulk that separate vacuum states with coefficients $c_k$ characterized by $c_{k+1}=(q/p)c_k$ (as in (\ref{eq:ansatz:nexcits})) is an eigenvector of the boundary operators $\widehat{B}_1^T$ and $B_L^T$ if one of two constraints below are satisfied:
\begin{equation}
 \label{eq:condition:nexcits:left}
 \begin{tabular}{|c|c|c|c|c|}
\hline
 Name & Condition & Eigenvalue $\Lambda$ & Value of $\tilde{c}_1$ & Value of $\tilde{c}_{n+1}$ \\
\hline
(A)&  $\dfrac{\alpha\beta}{\gamma\delta} \left(\dfrac{p}{q}\right)^{n'} e^s = 1$ & $\Lambda_\text{bulk}(\tilde{z}_1,\ldots,\tilde{z}_{n'})-(\alpha+\beta+\gamma+\delta)$ & $\tilde{c}_1 = -e^{-s}\gamma/\alpha$ & $\tilde{c}_{n+1}=-\beta/\delta $\\
\hline
(B) &$\left(\dfrac{p}{q}\right)^{n'} e^s = 1$ &  $\Lambda_\text{bulk}(\tilde{z}_1,\ldots,\tilde{z}_{n'})$ & $\tilde{c}_1=e^{-s}$ & $ \tilde{c}_{n+1}= 1$ \\
\hline
 \end{tabular}
\end{equation}
These constraints are exactly the same as in (\ref{eq:condition:nexcits}) except that $L-1-n$ is replaced by $n'$. Thus, if the first constraint of (\ref{eq:condition:nexcits}) is satisfied for given $n$ then it is possible to build:
\begin{enumerate}
 \item right eigenvectors with $n$ excitations of type (\ref{eq:excitv2}) separating vacuum states $\vide{c_k}{i}$ and with eigenvalue $\Lambda_\text{bulk}(z_1,\ldots,z_n)$ where the $z_k$'s satisfy the Bethe equations (\ref{eq:open:betheeqs}),
 \item left eigenvectors with $n'=L-1-n$ excitations of type (\ref{eq:excit:left}) separating vacuum states $\lvide{\tilde{c}_k}{i}$ and with eigenvalue $\Lambda_\text{bulk}(\tilde{z}_1,\ldots,\tilde{z}_{n'})-(\alpha+\beta+\gamma+\delta)$ where the $\tilde{z}_k$'s satisfy a second set of Bethe equations similar to the ones that correspond to the other part of the spectrum in \cite{degieressler2,degieressler3}.
\end{enumerate}
On the contrary, if the second constraint of (\ref{eq:condition:nexcits}) is satisfied, the correspondence to the two parts of the spectrum in \cite{degieressler2,degieressler3} is reversed. 

It is interesting to compare these results with the Matrix Ansatz (\ref{eq:matrixansatz}) for $s=0$. The construction of left and right eigenvectors described in this paper gives the \emph{complete} spectrum described in \cite{degieressler1,degieressler2,degieressler3} but only a subset of the right or the left eigenvectors. For $s=0$, it is always possible to choose $n=L-1$ within the constraint (B). The right eigenvectors describe the eigenvalues $\Lambda(z_1,\ldots,z_{L-1})-(\alpha+\beta+\gamma+\delta)$, which do not contain the ground state $\Lambda=0$ described by the matrix ansatz (\ref{eq:matrixansatz}). On the contrary, the construction explained in this paper give only one left eigenvector, which is a product state made of $\bra{0}+\bra{1}$, with eigenvalue $\Lambda=0$. It corresponds to the conservation of the total probability for $s=0$ and is thus the ground state. A short summary of these results is presented in table (\ref{tab:eigenvectors}). There is one known case \cite{esslerrittenberg,mallicksandow} for which the matrix ansatz does not give straighforwardly the stationary state and it corresponds precisely to the condition (A) in (\ref{eq:condition:nexcits}) for $s=0$: in this case, the coordinate Bethe ansatz corresponding to the constraint (A) in (\ref{eq:condition:nexcits}) gives it with all the $z_k$'s going to $1$. The eigenvectors presented here are never redundant with the Matrix Ansatz.

\begin{table}
\begin{center}
 \begin{tabular}{|c|c|c|}
\hline
 \multicolumn{3}{|c|}{Generic case $s\neq 0$} \\
\hline
  Part of the spectrum & Left eigenvect. & Right eigenvect. \\
\hline
  I (resp. II) &   CBA  & ? \\
\hline
 II (resp. I)  & ? & CBA \\
\hline
 \end{tabular}
 \begin{tabular}{|c|c|c|}
\hline
 \multicolumn{3}{|c|}{Special case $s=0$} \\
\hline
  Part of the spectrum & Left eigenvect. & Right eigenvect. \\
\hline
  $\Lambda=0$ &  $\bigotimes_{i=1}^L (\bra{0}+\bra{1})$  &  Matrix Ansatz \\
\hline
 Other $\Lambda$'s & ? & CBA \\
\hline
 \end{tabular}
\end{center}
 \caption{Summary of the eigenvectors described in this approach and the non-overlap with the Matrix Ansatz for $s=0$ (CBA=coordinate Bethe Ansatz described in this paper). In the first table, the part of the spectrum described by a set of eigenvectors depends on the constraint (\ref{eq:condition:nexcits}) that is satisfied. The second table is obtained by taking $n=L-1$ with constraint (B).\label{tab:eigenvectors}}
\end{table}

\section{Application: dynamics conditioned on the current in the case $n'=1$}
\label{sec:application}

The previous section shows how to build some of the left eigenvectors needed for the study of the dynamics $W_\text{cond}^{(j)}$ conditioned to produce a long time current $j$. Eq.~(\ref{eq:conditdynamics}) shows that the first left eigenvector of $\widehat{W}$ is needed. If this eigenvector belongs to the part I of the spectrum described in fig.~\ref{tab:eigenvectors}, then it is given by a coordinate Bethe Ansatz. For simplicity, only the case $s=-\ln(p/q)$, which corresponds to $n=L-2$ and $n'=1$, will be presented here: it is the first non-trivial case after $s=0$ since some of the left eigenvectors depend only on one Bethe root, noted $z$ from now on. The corresponding left eigenvectors are obtained from (\ref{eq:vide:left},\ref{eq:excit:left}) and are given by:
\begin{equation}
\begin{split}
 \bra{\psi} =&  \sum_{x=1}^L (Az^x+ A'z^{-x}) \left( \bigotimes_{i=1}^{x-1} \bra{0}+\tilde{c}\bra{1} \right)  \left(\sqrt{\frac{p}{q}}\right)^x\left( \bra{0}+\tilde{\nu}\bra{1} \right) \left( \bigotimes_{i=x+1}^{L}\bra{0}+\tilde{c}\frac{q}{p}\bra{1}\right)  \\
 & + A'' \left( \bigotimes_{i=1}^{L}\bra{0}+\tilde{c}\frac{q}{p}\bra{1}\right)
\end{split}
\end{equation}
where $\tilde{c}=p/q$ and $\tilde{\nu}$ is chosen in order to close the right boundary as in section \ref{subsec:ansatzvalidity}. The coefficients $A$, $A'$ and $A''$ are the ones given by the reflection condition at the left boundary in section \ref{subsec:firstreflection}. The Bethe root $z$ is a root of the polynomial equation
\begin{equation}
 z^{2L} \frac{V_1(z)V_L(z)}{V_1(1/z)V_L(1/z)}=1.
\end{equation}
It is not trivial to see that the first eigenvalue $\mu_1(s)$ of $\widehat{W}$ belongs to this part of the spectrum, even for $n'=1$. However numerical checks up to $L=6$ show that it is the case: it will be assumed that it is the case for general $L$ and $z$ is the corresponding solution of the previous equation. For a configuration
$\C=\tau_1\ldots\tau_L$ where $\tau_i=0$ or $1$, the component of the first eigenvectors of $\widehat{W}$ on $\ket{\C}$ is given by:
\begin{equation}
 \langle \mu_1(s) | \C \rangle = \sum_{x=1}^L (Az^x+ A'z^{-x}) \left( \prod_{i=1}^{x-1} \tilde{c}^{\tau_i} \right)  \left(\sqrt{\frac{p}{q}}\right)^x \tilde{\nu}^{\tau_x}  \left( \prod_{i=x+1}^{L} \left(\tilde{c}\frac{q}{p}\right)^{\tau_i}\right)   + A'' \left( \prod_{i=1}^{L} \left(\tilde{c}\frac{q}{p}\right)^{\tau_i}\right)
\end{equation}
This simple expression where all the numbers are known can be used directly to simulate the conditioned dynamics for the non-trivial value of $s=-\ln(p/q)$.

To illustrate the effect of the conditioning, one can look at the modified jumping rate of a single particle in the system. For a configuration $\C_y=0\ldots010\ldots0$ where the single particle is at position $y$, the particle can hop on the left side with probability $qe^{-s}U(y-1) /U(y)$ or to the right with probability $p e^{-s}U(y+1)/U(y)$. The contribution $U(y)=\langle \mu_1(s) | \C_{y}\rangle$ acts as an exterior potential and is given by:
\begin{equation}
 U(y) = \langle \mu_1(s) | \C_{y} \rangle = A'' \left(\tilde{c}\frac{q}{p}\right) + \sum_{x=1}^L (Az^x+ A'z^{-x})\left(\sqrt{\frac{p}{q}} \right)^x \left[ \tilde{c}\delta_{y<x} + \tilde{\nu}\delta_{xy} + \tilde{c}\frac{q}{p}\delta_{y>x}  \right]
\end{equation}
Similar computations with two particles in the system gives the interaction, which may not be short-range, between two particles in addition to this exterior potential. More detailed results about this interaction will be presented in a future work.

\section{Conclusion}

This paper shows how to construct eigenvectors of the transition matrix of the 
asymmetric exclusion process with different reservoirs at both boundaries modified to count the current, when the parameters satisfy some algebraic conditions. The Ansatz (\ref{eq:completeCBA}) shows that one must allow excitations to leave the system through the boundary reservoirs: the integrability condition takes the form of a compatibility condition (\ref{eq:compatibility:level2}) between the transmission coefficients of the $m$-th excitation and the reflexion coefficients of the $(m+1)$-th one.

A key feature in the identification of the special points specified by the constraints (\ref{eq:condition:nexcits}) is the existence of a free parameter $c$ in the local stationary measures $\vide{c}{i}$ that can be adjusted to the boundary operators $\widehat{B}_1$ and $B_L$. When $c$ is positive, it is related to the mean density of particles on the sites of the lattice. An excitation can be seen as a moving frontier between two local stationary measures with different parameters $c$. The detection of the special points where a coordinate Bethe Ansatz (\ref{eq:completeCBA}) is valid is then made easy once the relation between the left and parameters $c$ and $c'$ is known: one starts with $c_1$ fixed by the left boundary, computes $c_2$,\ldots,$c_{n+1}$ with the previous relation and tries to match it with the right reservoir. 

The second step consists in studying how excitations move in the bulk and interact, with the usual technique of the coordinate Bethe Ansatz. At the boundary, the determination of the reflection and transmission coefficients for the first excitation is easy and it is enough to get the Bethe equations if one already knows that the coordinate Bethe Ansatz is valid. Checking its validity, and thus the integrability, requires however to check that the transmission coefficient of the first excitation  are compatible with the reflection coefficient of the second one, etc. Once the Bethe equations are obtained, the ground state and the gap can be computed after some work \cite{degieressler1,degieressler2,degieressler3}.

The advantage of the method presented here is that it does not use special algebraic properties of the model and one may hope that it could be extended to other integrable models. Moreover, the approach followed here provides an easy way of finding the submanifold of the parameter space where this coordinate Bethe Ansatz is valid; however this restriction to a submanifold appears only on presence of two boundaries: most of the construction may be valid without restriction for a half-line with only one reservoir.

When the measure of the current is discarded by setting $s=0$, interesting relations with the Matrix ansatz appear. The only right eigenvector that is not given by the coordinate Bethe Ansatz is the one given by the matrix ansatz. When constraints (A) in (\ref{eq:condition:nexcits}) is satisfied, then the matrix ansatz is known to fail in this case (\cite{esslerrittenberg,mallicksandow}). Moreover, the decomposition of the eigenvectors into shocks and Bernoulli product measures is the one that appears when the matrix ansatz has finite representations. The parameter space contains submanifolds with a finite dimensional matrix ansatz (finite number of shocks) and submanifolds (\ref{eq:condition:nexcits}) with a finite number of Bethe excitations. From these considerations, the missing eigenvectors in the table \ref{tab:eigenvectors} may be related to a modified form of the matrix ansatz and it would be interesting to know how the two approaches are related. Coupled with (\ref{eq:condeigenvectors}), the knowledge of the missing eigenvectors would allow one to understand the full dynamics of ASEP conditioned to give a given current and may help to study the optimal density profile to produce this current.

One may be interested in extending this construction to other models. The Bethe roots $z_k$ are attached to the excitations in the eigenvectors and the spectrum is completely determined by them. The structure of the vacuum states does not play any role once the dynamics of their frontiers is known. It may be interesting to try to use other types of locally stationary measures as vacuum states in other models and check if their boundaries have integrable motions.

A challenge is the determination of the structure of the eigenvectors for generic value of the parameters. The works \cite{baseilhackoizumi,galleas} are first steps in this direction for the XXZ spin chain. It would be interesting to find a physical interpretation in terms of shocks and Bernoulli product measures of these results in the case of the ASEP. It would also be very useful to see if the parametrization of the spectrum done \cite{galleas} in terms of Bethe roots also has a structure that generalizes the one presented here.

Finally, it would be interesting to extend this formalism to other fields of statistical physics. Two active domains where integrability plays an important role and for which boundary conditions are relevant are tilings and loop models. The knowledge of the eigenvectors and of the nature of the excitations in this models may provide interesting information about the typical configurations of these other models.

\acknowledgements
I would like to thank Bernard Derrida for having proposed me to work on this problem and for interesting discussions. I am also grateful to Gunter Sch\"utz, F.H. Jafarpour and Tomohiro Sasamoto for fruitful discussions. I would also like to thank the anonymous referee for his interesting and relevant comments and suggestions. This work was supported by a post-doctoral research fellowship of the Humboldt foundation.

\bibliographystyle{unsrt}
%\bibliography{opencba2}

\end{document}